\documentclass[aps,prd,twocolumn,superscriptaddress,showkeys,floatfix,nofootinbib]{revtex4-2}

\usepackage{graphicx}
\usepackage{amssymb}
\usepackage{amsmath}
\usepackage{bm}

\begin{document}

\title{Photon localization: a comparative study}


\author{Laura M.~Ferguson}
\email{itw023@mail.usask.ca}
\affiliation{Department of Physics and Engineering Physics, 
University of Saskatchewan, 116 Science Place, Saskatoon, Canada SK, S7N 5E2}
\affiliation{Department of Physics, Durham University, Durham, United Kingdom}

\author{Ricardo Rojas Castellano}
\email{awf345@mail.usask.ca}
\affiliation{Department of Physics and Engineering Physics, 
University of Saskatchewan, 116 Science Place, Saskatoon, Canada SK, S7N 5E2}
\affiliation{Departamento de Investigaci\'on en F\'isica,
  Universidad de Sonora, Hermosillo, Sonora, Mexico}

\author{Kayla Oltman}
\email{kao643@mail.usask.ca}
\affiliation{Department of Physics and Engineering Physics, 
University of Saskatchewan, 116 Science Place, Saskatoon, Canada SK, S7N 5E2}

\author{Joshua G.~Fenwick}
\email{joshua.fenwick@usask.ca}
\affiliation{Department of Physics and Engineering Physics, 
University of Saskatchewan, 116 Science Place, Saskatoon, Canada SK, S7N 5E2}


\author{Rainer Dick}
\email{rainer.dick@usask.ca}
\affiliation{Department of Physics and Engineering Physics, 
University of Saskatchewan, 116 Science Place, Saskatoon, Canada SK, S7N 5E2}



\begin{abstract} 
  We compare different measures for photon localization in terms of two-dimensional
  Gaussian wave
  packets $\bm{\phi}(\bm{k},t)=\phi(\bm{k},t)\bm{\epsilon}(\bm{k})$
  of width $\Delta k$.
  We find that all measures start to coalesce if the wave packet has evolved for times
  $t$ which are larger than a few $(c\Delta k)^{-1}$.
  However, the Landau-Peierls wave function $\bm{\phi}(\bm{x},t)$ 
  yields the largest initial offset $\langle r\rangle-ct>0$ while the Fourier
  transform $\psi(\bm{x},t)$ of $\phi(\bm{k},t)$ yields the
  smallest offset $\langle r\rangle-ct>0$ and converges
  towards $\langle r\rangle-ct=0$ fastest.\\
  We also discuss local detection of photons through a model detector consisting of ions
  in ion traps. The position-dependent detection
  probabilities are inferred from the scattering matrix.
  We find that the local detection probability for photons
  can be expressed in terms of three of the proposed localization measures,
  viz.~the Landau-Peierls wave function, the energy wave function,
  and the Hawton density. Those three proposals
  also remain close throughout the time evolution of the single-photon wave packet.
  However, the energy wave function is locally related
  to electromagnetic fields and is
  therefore confirmed as the preferred tool for the description
  of local photon densities.\\
  The novelty of our observations concerns three items, viz.~the possibility to express
  local photon absorption matrix elements through
  three of the proposed localization measures, the coalescence
  of all the proposed localization measures after sufficiently long evolution
  of a photon wave packet, and the striking differences of some of the
  proposed localization measures during early wave packet evolution.
\end{abstract}


\maketitle



\section{Introduction}\label{sec:intro}

Photons provide a basic ingredient for technical realizations
of quantum cryptography and quantum communication, and
they also provide a promising platform for quantum computing.
Technical implementations of photonic techniques for quantum computing
will require nanoscale integration of photon sources and detectors
\cite{ref:nano0,ref:nano1,ref:nano2,ref:nano3},
thus also implying a high spatial resolution of photon detection.
Single-photon detection with a resolution
of 12.6 microns has been demonstrated in superconducting nanowire
single-photon detectors (SNSPDs) \cite{ref:kong}.
Since the superconducting energy gap in the MoSi films
used in these devices is around $1\,\mathrm{meV}$, far infrared photons 
up to millimeter wavelengths could be
detected by these devices.
Furthermore, nanoscale avalanche photodiodes (nanoAPDs) have achieved 
spatial resolution of $250\,\mathrm{nm}$ for detection of photons with
wavelength $\lambda=488\,\mathrm{nm}$ \cite{ref:hayden}.
The high position resolution $\Delta x$ in SNSPDs and nanoAPDs is achieved through
the size of the absorbing region\footnote{In SNSPDs, the readout technique
is also important for identifying the segment of the nanowire where the photon
was absorbed \cite{ref:kong}.}, while the photon detection
event as such involves absorption across an intrinsic energy gap $E_g$
in the device, viz.~the superconducting energy gap in SNSPDs or
the band gap of the n-doped component in a nanoAPD.

High local resolution of photon detection and nanophotonic integration
makes it worthwhile to revisit the longstanding issue of the proper theoretical
description of photon location: Given a photon wave packet $\bm{\phi}(\bm{k},t)$
in momentum space, what is the proper (or at least, suitable) density for
the probability $P_V(t)$ that the photon will trigger a signal in the volume $V$
at time $t$? We will consider photon absorption in ion traps
to study the question what the position-dependence of
photon absorption can tell us about photon localization measures.

We would like to emphasize that the question for ``location'' is
not ``where is an energy-momentum quantum located at time $t$?'', but rather:
\textit{Where does an energy-momentum quantum have a high probability to trigger
  a signal at time $t$?} This is exactly the same for electrons, photons, or any
other energy-momentum quanta. The difference between electrons and photons is that
electrons are not absorbed upon observation and can therefore trigger a continuous
sequence of signals, e.g., a trajectory in a particle detector. However, the
basic question for location only has operational meaning in terms of detection.
Therefore, we suggest that a photon position probability
density $\varrho_\gamma(\bm{x},t)$ should satisfy the following requirement:
Photon detection rates by a detector located in a position $\bm{a}_I$ should be closely
correlated with the values of the photon probability density $\varrho_\gamma(\bm{x},t)$
near the detector site $\bm{a}_I$. Ideally, this would mean that the detection rate
is proportional to $\varrho_\gamma(\bm{a}_I,t)$. However, we will see in
Sec.~\ref{sec:detect} that this is asking too much. Instead, we will find
that the local photon detection rate can be expressed in terms of
sampling the values of the wave functions 
corresponding to $\varrho_\gamma(\bm{x},t)$ near the detector site $\bm{a}_I$.
Specifically, we will find that our
criterion supports the Landau-Peierls wave function $\bm{\phi}(\bm{x},t)$,
the energy
wave function $\bm{\Psi}(\bm{x},t)=-\,\mathrm{i}\sqrt{2\epsilon_0}\bm{E}^{(+)}(\bm{x},t)$,
and the factors that contribute to the Hawton density $\varrho_H(\bm{x},t)$.

For the introduction of the major contemporary proposals for photon wave functions
and position probability densities, we recall that 
a single-photon wave packet in $\bm{k}$ space involves two polarization
components $\phi_\alpha(\bm{k},t)=\phi_\alpha(\bm{k})\exp(-\,\mathrm{i}ckt)$
corresponding to the polarization vectors $\bm{\epsilon}_\alpha(\bm{k})$,
and a corresponding vector wave packet
\begin{equation}
  \bm{\phi}(\bm{k},t)=\sum_{\alpha=1}^2\phi_\alpha(\bm{k},t)\bm{\epsilon}_\alpha(\bm{k}).
\end{equation}
Here, we define the wave packet such that the normalization condition is
\begin{equation}
\int\!d^3\bm{k}\left|\bm{\phi}(\bm{k},t)\right|^2=1,
\end{equation}
and energy and momentum are
\begin{equation}\label{eq:Epacket}
  E=\int\!d^3\bm{k}\,\hbar ck\left|\bm{\phi}(\bm{k},t)\right|^2,
\end{equation}
and
\begin{equation}\label{eq:Ppacket}
  \bm{p}=\int\!d^3\bm{k}\,\hbar\bm{k}\left|\bm{\phi}(\bm{k},t)\right|^2,
\end{equation}
respectively.
The corresponding vector potential is
\begin{eqnarray}
  \bm{A}(\bm{x},t)&=&\bm{A}^{(+)}(\bm{x},t)+\bm{A}^{(-)}(\bm{x},t)
\end{eqnarray}
with the positive frequency component $\bm{A}^{(+)}(\bm{x},t)$
and the negative frequency component $\bm{A}^{(-)}(\bm{x},t)$,
  \begin{eqnarray}\nonumber
    \bm{A}^{(+)}(\bm{x},t)&=&[\bm{A}^{(-)}(\bm{x},t)]^+
    \\ \label{eq:A+1}
    &=&\sqrt{\frac{\hbar\mu_0 c}{8\pi^3}}\int\!\frac{d^3\bm{k}}{\sqrt{2k}}\,
    \bm{\phi}(\bm{k},t)\exp(\mathrm{i}\bm{k}\cdot\bm{x}).
  \end{eqnarray}
  This yields the corresponding components of the electromagnetic fields\footnote{
  We work in Coulomb gauge and use
  $\bm{E}(\bm{x},t)\equiv\bm{E}_\perp(\bm{x},t)=-\,\partial\bm{A}(\bm{x},t)/\partial t$,
  since
  $\bm{E}_\|(\bm{x},t)=-\,c\bm{\nabla}A^0(\bm{x},t)$ is given in terms of the charged
  particle fields and yields the Coulomb terms in Coulomb gauge.}
  \begin{equation}
    \bm{E}^{(\pm)}(\bm{x},t)=-\,\frac{\partial}{\partial t}\bm{A}^{(\pm)}(\bm{x},t),
  \end{equation}
  \begin{equation}
    \bm{B}^{(\pm)}(\bm{x},t)=\bm{\nabla}\times\bm{A}^{(\pm)}(\bm{x},t).
  \end{equation}

Proposals for probability densities for photon position involve
the normalized electromagnetic energy density $\mathcal{H}(\bm{x},t)/E$
either in the classical form \cite{ref:bialynicki1994}
\begin{equation}\label{eq:Hc}
  \mathcal{H}_c(\bm{x},t)=\frac{\epsilon_0}{2}\bm{E}^2(\bm{x},t)
  +\frac{1}{2\mu_0}\bm{B}^2(\bm{x},t),
\end{equation}
or in the form that does not include the fast oscillating,
short-wavelength interference terms,
\begin{eqnarray} \nonumber
  \mathcal{H}(\bm{x},t)&=&
  \epsilon_0\bm{E}^{(-)}(\bm{x},t)\cdot\bm{E}^{(+)}(\bm{x},t)
  \\ \label{eq:H1}
  &&+\,\frac{1}{\mu_0}\bm{B}^{(-)}(\bm{x},t)\cdot\bm{B}^{(+)}(\bm{x},t).
\end{eqnarray}
 
The transition from (\ref{eq:Hc}) to (\ref{eq:H1}) could be
understood as a consequence of spatial coarse graining 
of $\mathcal{H}_c(\bm{x},t)$. If electromagnetic fields are dominated
by contributions from a particular wave vector $\bm{k}$, $\Delta k\ll|\bm{k}|$,
then integration over a volume $V>(2\pi/|\bm{k}|)^3$ eliminates the fast
oscillating components in $\mathcal{H}_c(\bm{x},t)$,
\begin{equation}\label{eq:coarse1}
  \int_V\!d^3\bm{x}\,\mathcal{H}_c(\bm{x},t)
  \to\int_V\!d^3\bm{x}\,\mathcal{H}(\bm{x},t).
\end{equation}
However, this assumption of spatial coarse graining to eliminate
terms like $[\bm{E}^{(\pm)}(\bm{x},t)]^2$
appears rather ad hoc.
Based on our operational definition of position probability density as
being defined by local detection probability, we expect that
only products of
the form $\bm{E}^{(-)}(\bm{x},t)\otimes\bm{E}^{(+)}(\bm{x},t)$
or $\bm{B}^{(-)}(\bm{x},t)\otimes\bm{B}^{(+)}(\bm{x},t)$
can play a role in the definition of photon probability densities:
Photon absorption matrix elements $\mathcal{M}_{f,i\gamma}$ can only
contain positive frequency components,
whence only products of negative frequency components with positive frequency
components can appear in $|\mathcal{M}_{f,i\gamma}|^2$.
Therefore, we do not need any coarse graining of the form (\ref{eq:coarse1})
to justify localization measures involving only products of negative frequency
components with positive frequency components.


Another proposal for a photon position probability density
involves the normalized ``energy wave function'' $\bm{\Psi}(\bm{x},t)/\sqrt{E}$
\cite{ref:MW,ref:sipe,ref:bialynicki1996,ref:SR},
\begin{equation}\label{eq:Psi1}
  \bm{\Psi}(\bm{x},t)=-\,\mathrm{i}\sqrt{2\epsilon_0}\bm{E}^{(+)}(\bm{x},t).
\end{equation}
The corresponding photon position probability density $|\bm{\Psi}(\bm{x},t)|^2/E$
can be motivated by Glauber's observation that the local photon detection rate from
a state $|\gamma\rangle$ can be approximated through the expectation value
$\langle\gamma|\bm{\mathrm{E}}^{(-)}(\bm{x},t)\bm{\mathrm{E}}^{(+)}(\bm{x},t)|\gamma\rangle$
\cite{ref:glauber1965}. Here, we use upright notation to denote the electric field
operators.

The designation ``energy wave function'' for $\bm{\Psi}(\bm{x},t)$
can be motivated from the fact
that for narrow photon wave packets in $\bm{k}$ space,
$\Delta k\ll|\langle\bm{k}\rangle|$, we have
$|\bm{\Psi}(\bm{x},t)|^2\simeq\mathcal{H}(\bm{x},t)$, where $\mathcal{H}(\bm{x},t)$
is the energy density (\ref{eq:H1}).
Similar to the relation between $\mathcal{H}_c(\bm{x},t)$
and $\mathcal{H}(\bm{x},t)$, $|\bm{\Psi}(\bm{x},t)|^2$ could be inferred from
spatial coarse graining of $\epsilon_0\bm{E}^2(\bm{x},t)$. However, note
that \textit{the energy wave functions as a measure for photon probability density
  does not require spatial coarse graining for its definition}, but is directly
inferred and motivated by Glauber's
calculation of photon absorption rates.

All three of the principally different
densities (\ref{eq:Hc},\ref{eq:H1},\ref{eq:Psi1})
yield the energy (\ref{eq:Epacket}) upon spatial integration,
\begin{eqnarray} \nonumber
  E&=&\int\!d^3\bm{x}\,|\bm{\Psi}(\bm{x},t)|^2=\int\!d^3\bm{x}\,\mathcal{H}_c(\bm{x},t)
  \\
  &=&\int\!d^3\bm{x}\,\mathcal{H}(\bm{x},t).
\end{eqnarray}

 Further proposals for photon position probability densities involve
the Landau-Peierls vector wave function\footnote{We include Refs.~\cite{ref:mandel,ref:MW}
with the Landau-Peierls wave function because this wave function would also emerge
as an eigenfunction through coherent states of Mandel's photon detection operator
\begin{equation}\label{eq:mandelD}
  \bm{a}(\bm{x},t)=\frac{1}{\sqrt{2\pi}^3}\int\!d^3\bm{k}\,
  \exp[\mathrm{i}(\bm{k}\cdot\bm{x}-ckt)] \bm{a}(\bm{k}),
\end{equation}
\begin{equation}
  \bm{a}(\bm{k})=\sum_{\alpha=1}^2a_\alpha(\bm{k})\bm{\epsilon}_\alpha(\bm{k}),
\end{equation}
with commutators
\begin{equation}
  [a_i(\bm{k}),a_j^+(\bm{k}')]=\left(\delta_{ij}-\frac{k_ik_j}{\bm{k}^2}\right)
  \delta(\bm{k}-\bm{k}').
\end{equation}
Mandel's detection operator is the interaction picture version
of the operator (\ref{eq:mandelS}).}
\cite{ref:mandel,ref:MW,ref:LP,ref:bialynicki1996},
\begin{equation}\label{eq:LP}
  \bm{\phi}(\bm{x},t)=\frac{1}{\sqrt{2\pi}^3}\int\!d^3\bm{k}\,\bm{\phi}(\bm{k},t)
  \exp(\mathrm{i}\bm{k}\cdot\bm{x}),
\end{equation}
\begin{equation}
  \bm{\nabla}\cdot\bm{\phi}(\bm{x},t)=0,
\end{equation}
and 
Cook's 2-component vector wave function \cite{ref:cook1,ref:cook2},
\begin{equation}\label{eq:cook1}
  \bm{\phi}_1(\bm{x},t)=\frac{1}{\sqrt{2}}\,\bm{\phi}(\bm{x},t),
  \end{equation}
\begin{equation}\label{eq:cook2}
  \bm{\phi}_2(\bm{x},t)=\frac{1}{4\sqrt{\pi}^3}\int\!d^3\bm{k}\,\hat{\bm{k}}\times
  \bm{\phi}(\bm{k},t)\exp(\mathrm{i}\bm{k}\cdot\bm{x}),
  \end{equation}
\begin{equation}
  \bm{\nabla}\cdot\bm{\phi}_a(\bm{x},t)=0.
\end{equation}

The Landau-Peierls wave function (\ref{eq:LP}) has been 
criticized for its nonlocal relation to electromagnetic fields and
energy-momentum densities, but it yields a suggestive
position representation of the photon state in the form
\begin{eqnarray}\nonumber
  |\phi(t)\rangle&=&\int\!d^3\bm{k}\,\bm{a}^+(\bm{k})|0\rangle\cdot
  \bm{\phi}(\bm{k})\exp(-\,\mathrm{i}ckt)
  \\ \label{eq:LPmandel}
  &=&\int\!d^3\bm{x}\,\bm{a}^+(\bm{x})|0\rangle\cdot
  \bm{\phi}(\bm{x},t),
\end{eqnarray}
where
\begin{equation}\label{eq:mandelS}
  \bm{a}(\bm{x})=\frac{1}{\sqrt{2\pi}^3}\int\!d^3\bm{k}\,\bm{a}(\bm{k})
  \exp(\mathrm{i}\bm{k}\cdot\bm{x}),
\end{equation}
\begin{equation}
[a_i(\bm{x}),a_j^+(\bm{x}')]=\delta_{ij}^\perp(\bm{x}-\bm{x}'),
\end{equation}
is the Schr\"odinger picture version of Mandel's photon detection
operator (\ref{eq:mandelD}).
Furthermore, we will see in Sec.~\ref{sec:packet} that
the Landau-Peierls density $|\bm{\phi}(\bm{x},t)|^2$
closely tracks $|\bm{\Psi}(\bm{x},t)|^2/E=
2\epsilon_0\bm{E}^{(-)}(\bm{x},t)\cdot\bm{E}^{(+)}(\bm{x},t)/E$ throughout
the evolution of a single-photon wave packet. 

The Landau-Peierls wave function can be calculated from the energy wave function
through \cite{ref:MW,ref:bialynicki1996}
\begin{equation}\label{eq:Psitophi3d}
  \bm{\phi}(\bm{x},t)=\frac{1}{4\sqrt{2\pi^3}\sqrt{\hbar c}}
  \int\!d^3\bm{x}'\,\frac{\bm{\Psi}(\bm{x}',t)}{|\bm{x}-\bm{x}'|^{5/2}},
\end{equation}
see Appendix A.

Formal inversion of Eq.~(\ref{eq:Psitophi3d}) motivates a relation
that involves a singular kernel in three dimensions,
\begin{equation}\label{eq:phitoPsi3d}
  \bm{\Psi}(\bm{x},t)\sim\frac{3\sqrt{\hbar c}}{8\sqrt{2\pi^3}}
  \int\!d^3\bm{x}'\,\frac{\bm{\phi}(\bm{x}',t)}{|\bm{x}-\bm{x}'|^{7/2}}.
\end{equation}

Representing $\bm{\Psi}(\bm{x},t)$ on the basis of (\ref{eq:phitoPsi3d}) has occasionally
been used in the literature, but should
be avoided because it presumes exchange of $\bm{k}$-space and $\bm{x}'$-space
integrals for powers of $k$ where this is not possible, i.e.,
Eq.~(\ref{eq:phitoPsi3d}) follows from a formal extension of the methods that lead to
Eq.~(\ref{eq:Psitophi3d}) into a parameter range where the derivation does not
apply anymore, see Appendix A.
Indeed, the fact that the Landau-Peierls wave function can be calculated from the
energy wave function through Eq.~(\ref{eq:Psitophi3d}),
whereas the reverse Eq.~(\ref{eq:phitoPsi3d}) is not true,
contributed to Bia{\l}ynicki-Birula's
preference for the energy wave function as the proper candidate for a photon
wave function \cite{ref:bialynicki1996}.

In a different approach, Hawton had proposed to base the photon current density on
$j^\nu=-\,\mathrm{i}\epsilon_0 c A_\mu F^{\mu\nu}/\hbar$ \cite{ref:HawtonMelde,ref:Hawton}. 
The density $j^0/c$ yields after spatial coarse graining a density
\begin{eqnarray} \nonumber
  \varrho_H(\bm{x},t)&=&\mathrm{i}\,\frac{\epsilon_0}{\hbar}\left[
    \bm{E}^{(-)}(\bm{x},t)\cdot\bm{A}^{(+)}(\bm{x},t)\right.
    \\ \label{eq:rhoH}
    &&-\left.\bm{A}^{(-)}(\bm{x},t)\cdot\bm{E}^{(+)}(\bm{x},t)\right].
\end{eqnarray}
As is the case for the energy wave function,
we will see in the calculation of photon absorption 
that $\varrho_H(\bm{x},t)$ can also be inferred directly from the
corresponding absorption matrix elements without recourse to
spatial coarse graining of any classical density.

The Hawton density can be expressed in terms of the energy
wave function (\ref{eq:Psi1}) in the form
\begin{eqnarray}\label{eq:rhoHpsi}
\varrho_H(\bm{x},t)&=&\int\!d^3\bm{x}'\,
\frac{1}{4\pi^2\hbar c|\bm{x}-\bm{x}'|^2}
\\ \nonumber
&&\times\left[\bm{\Psi}^+(\bm{x},t)\cdot\bm{\Psi}(\bm{x}',t)
+\bm{\Psi}^+(\bm{x}',t)\cdot\bm{\Psi}(\bm{x},t)\right],
\end{eqnarray}
see Appendix A, especially Eqs.~(\ref{eq:Psi2A})
and (\ref{eq:rhoHpsi2}), for the derivation.
Eq.~(\ref{eq:rhoHpsi}) helps us to understand
why the not manifestly positive definite Hawton density yields
positive definite results when the energy wave function
has single maxima for
$[\mathrm{Re}\,\bm{\Psi}(\bm{x},t)]^2$ and $[\mathrm{Im}\,\bm{\Psi}(\bm{x},t)]^2$.
The integral in (\ref{eq:rhoHpsi}) will be dominated by the positive
contributions from the regions near the maxima.

We will also encounter a corresponding Hawton tensor,
\begin{eqnarray} \nonumber
  \underline{\varrho}_H(\bm{x},t)&=&\mathrm{i}\,\frac{\epsilon_0}{\hbar}\left[
    \bm{E}^{(-)}(\bm{x},t)\otimes\bm{A}^{(+)}(\bm{x},t)\right.
    \\ \label{eq:rhoHt}
    &&-\left.\bm{A}^{(-)}(\bm{x},t)\otimes\bm{E}^{(+)}(\bm{x},t)\right],
\end{eqnarray}
$\varrho_H(\bm{x},t)=\mathrm{tr}\,\underline{\varrho}_H(\bm{x},t)$,
in the calculation of photon absorption rates.

Finally, investigations of relativistic probability densities for Klein-Gordon
and Dirac fields \cite{ref:FD2023} and Hawton's recent investigations of position
observables in photon quantum mechanics 
also motivate consideration of
the Fourier transformations of the polarization
amplitudes $\phi_\alpha(\bm{k},t)=\phi_\alpha(\bm{k})\exp(-\,\mathrm{i}ckt)$ into direct space,
\begin{equation}\label{eq:FD}
  \psi_\alpha(\bm{x},t)=\frac{1}{\sqrt{2\pi}^3}\int\!d^3\bm{k}\,\phi_\alpha(\bm{k},t)
  \exp(\mathrm{i}\bm{k}\cdot\bm{x}).
\end{equation}
Hawton now considers $\sum_\alpha|\psi_\alpha(\bm{x},t)|^2$ as the preferred
candidate for a photon position probability density \cite{ref:Hawton2}.

Eq.~(\ref{eq:FD}) yields a photon position probability density 
that is different from the Landau-Peierls
density, $\sum_\alpha|\psi_\alpha(\bm{x},t)|^2\neq |\bm{\phi}(\bm{x},t)|^2$, but it shares
the same advantages of being automatically normalized and ensuring a standard position-momentum
uncertainty relation, $\Delta x\Delta k_x\ge 1/2$, and it
also yields a suggestive position representation
of the photon state in the form
\begin{eqnarray}\nonumber
  |\phi(t)\rangle&=&\int\!d^3\bm{k}\sum_{\alpha=1}^2 a_\alpha^+(\bm{k})|0\rangle
  \phi_\alpha(\bm{k})\exp(-\,\mathrm{i}ckt)
  \\ \label{eq:gamma2}
  &=&\int\!d^3\bm{x}\sum_{\alpha=1}^2 a_\alpha^+(\bm{x})|0\rangle
  \psi_\alpha(\bm{x},t).
\end{eqnarray}
The detection operator
\begin{equation}
  a_\alpha(\bm{x})=\frac{1}{\sqrt{2\pi}^3}\int\!d^3\bm{k}\,a_\alpha(\bm{k})
  \exp(\mathrm{i}\bm{k}\cdot\bm{x}),
\end{equation}
\begin{equation}
[a_\alpha(\bm{x}),a_\beta^+(\bm{x}')]=\delta_{\alpha,\beta}\delta(\bm{x}-\bm{x}'),
\end{equation}
is nonlocally related to Mandel's detection operator,
\begin{eqnarray}\nonumber
  a_\alpha(\bm{x})&=&\frac{1}{(2\pi)^3}\int\!d^3\bm{x}'\int\!d^3\bm{k}\,
  \exp[\mathrm{i}\bm{k}\cdot(\bm{x}-\bm{x}')]
  \\ \label{eq:mandelmandel}
  &&\times\bm{\epsilon}_\alpha(\bm{k})\cdot
  \bm{a}(\bm{x}').
\end{eqnarray}

On the other hand, $\psi_\alpha(\bm{x},t)$ shares the disadvantage
of $\bm{\phi}(\bm{x},t)$ of being nonlocally related to the electromagnetic fields
that correspond to the photon wave packet $\bm{\phi}(\bm{k},t)$,
thus seemingly implying a nonlocal interaction between photons and charged particles
if either $\sum_\alpha|\psi_\alpha(\bm{x},t)|^2$ or $|\bm{\phi}(\bm{x},t)|^2$ are
considered as local photon densities. However, we take an agnostic approach and
simply ask what the different possible proposals for local photon detection probability
densities entail for a single-photon wave packet $\bm{\phi}(\bm{k},t)$.

Both of the representations (\ref{eq:LPmandel}) and (\ref{eq:gamma2}) are
correct and mutually consistent representations of the single-photon
state $|\phi(t)\rangle$. Both of those representations also resemble a
familiar equation in Schr\"odinger field theory for
single-electron states with spinor
$\chi(\bm{x},t)=\left(\chi_{1/2}(\bm{x},t),\chi_{-\,1/2}(\bm{x},t)\right)$,
\begin{equation}\label{eq:electron1}
  |\chi(t)\rangle=\int\!d^3\bm{x}\sum_{s=-\,1/2}^{1/2}
  c_s^+(\bm{x})|0\rangle\chi_s(\bm{x},t),
\end{equation}
where the relevant anticommutator of the electron operators is
\begin{equation}
  \{c_s(\bm{x}),c_{s'}^+(\bm{x}')\}=\delta_{s,s'}\delta(\bm{x}-\bm{x}').
\end{equation}
In the spirit of the Born interpretation, we interprete Eq.~(\ref{eq:electron1})
to mean that an electron with spin polarization $s$
can be found in the point $\bm{x}$ at time $t$ with
a probability density amplitude $\chi_s(\bm{x},t)$.
\textit{However, such an interpretation cannot be simultaneously correct for
  both of the correct representations (\ref{eq:LPmandel}) and (\ref{eq:gamma2})
  for the single-photon state $|\phi(t)\rangle$.} The reason is that the
nonlocal relation (\ref{eq:mandelmandel}) also holds for the single-photon
amplitudes,
\begin{eqnarray}\nonumber
  \psi_\alpha(\bm{x},t)&=&\frac{1}{(2\pi)^3}\int\!d^3\bm{x}'\int\!d^3\bm{k}\,
  \exp[\mathrm{i}\bm{k}\cdot(\bm{x}-\bm{x}')]
  \\ \label{eq:mandelmandel2}
  &&\times\bm{\epsilon}_\alpha(\bm{k})\cdot
  \bm{\phi}(\bm{x}',t),
\end{eqnarray}
and we will see in Sec.~\ref{sec:packet} that $\sum_\alpha|\psi_\alpha(\bm{x},t)|^2$
can be very different from $|\bm{\phi}(\bm{x},t)|^2$.
This cautions us that neither of the two
expressions $\sum_\alpha|\psi_\alpha(\bm{x},t)|^2$ or $|\bm{\phi}(\bm{x},t)|^2$
may be a probability density for photon detection. However, the detector
analysis in Sec.~\ref{sec:detect} makes
$|\bm{\phi}(\bm{x},t)|^2$
appear as a stronger candidate than $\sum_\alpha|\psi_\alpha(\bm{x},t)|^2$.

In Sec.~\ref{sec:detect}, we will revisit calculations of local photon detection rates
through absorption through atomic states, which according to conventional
wisdom favor the energy wave function (\ref{eq:Psi1}). However, a thorough reevaluation
shows that detector analyis alone
cannot uniquely identify a ``correct'' measure for the photon position
probability density.
We will find that three proposals for local photon density, viz.~the
Landau-Peierls wave function, the energy wave function, and the Hawton density,
can provide local descriptions for photon detection.

Following the observations of Sec.~\ref{sec:detect}, 
we turn to a study of the behavior of the
different proposals for photon probability density through their implications for
a two-dimensional Gaussian single-photon packet in $\bm{k}$ space.
This shows that $\sum_\alpha|\psi_\alpha(\bm{x},t)|^2$ and $\mathcal{H}(\bm{x},t)/E$
behave very differently from the Landau-Peierls wave function, the energy wave function,
and the Hawton density during early evolution of the wave packet, while all
proposed photon densities
start to coalesce towards the outgoing
spherical wave $\langle r\rangle-ct=0$ for late times.
This is described in Sec.~\ref{sec:packet}.
We summarize our findings in Sec.~\ref{sec:conc}. Appendix A reviews the different
nonlocal relations between the energy wave function, the vector potential, the
Landau-Peierls wave function,
and the Hawton density $\varrho_H(\bm{x},t)$ through convolutions.
Appendix B derives the matrix elements for photon absorption in an array
of ion traps from integration of the second-quantized evolution equation.

\section{Local photon detection rates: absorption in ions}\label{sec:detect}

We introduce an ion trap model as a
model system for the study of the position-dependence of photon absorption rates.
The virtue of 
our model is its relative simplicity for the exploration
of position-dependence of photon absorption, to inform which of the proposed
photon localization measures comply with the operational definition set out in
Sec.~\ref{sec:intro}.
However, this is not meant to model nanoAPDs as the highest spatial resolution
detection devices, since the minimal spatial resolution of ion traps will be limited
by the ion position uncertainty in the trap. The model is chosen for maximal
simplicity for studying the problem at hand: What do basic photon detection
mechanisms tell us about proposed probability densities for photon position?

With the normalization
\begin{equation}\label{eq:defMfi}
  S_{f|i\gamma}=-\,\mathrm{i}\mathcal{M}_{f|i\gamma}\delta(\omega_{f,i}-ck)
\end{equation}
between the scattering matrix element $S_{f|i\gamma}$ and the scattering amplitude
$\mathcal{M}_{f|i\gamma}$ for absorption of a photon $\gamma$ through a
transition $|i\rangle\to|f\rangle$ between electronic states,
the differential absorption rate for photons
with frequency $ck=\omega_{f,i}=(E_f-E_i)/\hbar$ is
\begin{equation}\label{eq:dGamma1}
  \frac{d\Gamma_{i\gamma\to f}}{d\Omega}
  =\frac{\omega_{f,i}^2}{2\pi c^3}\left|\mathcal{M}_{f|i\gamma}\right|^2.
\end{equation}
Therefore, for the question what local absorption rates
tell us about probability densities for photon signals, we are particularly
interested in the position dependence of the scattering matrix elements
$S_{fi}$, or the corresponding scattering amplitudes $\mathcal{M}_{fi}$, which
are gotten after evaluation of the second-quantized operators.

Eqs.~(\ref{eq:defMfi},\ref{eq:dGamma1}) hold in this form for absorption
of monochromatic photon states $|\bm{k},\alpha\rangle=a_\alpha^+(\bm{k})|0\rangle$.
The corresponding first-order scattering matrix elements for photon absorption
due to transitions $|i\rangle\to |f\rangle$
between electronic states are
\begin{eqnarray}\nonumber
  S_{fi}&=&\sqrt{\alpha_S c\omega_{f,i}}\,\bm{\epsilon}_\alpha(\bm{k})\cdot
    \langle f|\bm{\mathrm{x}}\exp(\mathrm{i}\bm{k}\cdot\bm{\mathrm{x}})|i\rangle
    \delta(\omega_{f,i}-ck)
    \\ \label{eq:matrix1}
    &\simeq&\sqrt{\alpha_S c\omega_{f,i}}\,\bm{\epsilon}_\alpha(\bm{k})\cdot
      \langle f|\bm{\mathrm{x}}|i\rangle\delta(\omega_{f,i}-ck),
\end{eqnarray}
where $\alpha_S$ is the fine structure constant
and we use upright notation, $\bm{\mathrm{x}}$, for the position operator.
Dipole approximation is used in the last equation.

One might expect that Glauber's calculation of local photon detection
rates through the dipole interaction term \cite{ref:glauber1965}
settles the question for the proper photon position probability density
in favor of the energy wave function (\ref{eq:Psi1}), which is also favored
by Sipe \cite{ref:sipe}, Bia{\l}ynicki-Birula \cite{ref:bialynicki1996}, 
and Smith and Raymer \cite{ref:SR}. However, other leading experts on the topic
disagree. Mandel's emphasis of the photon detection operator (\ref{eq:mandelD})
\cite{ref:mandel,ref:MW}
amounts to support for the Landau-Peierls wave function (\ref{eq:LP}),
whereas Hawton had emphasized the attractive features of the
density $\varrho_H(\bm{x},t)$ (\ref{eq:rhoH}) in Refs.~\cite{ref:HawtonMelde,ref:Hawton}
and more recently favors $\sum_\alpha|\psi_\alpha(\bm{x},t)|^2$ \cite{ref:Hawton2}.
Indeed, we will see that local photon detection rates can be expressed by three of the
proposed measures. Therefore, while Glauber's result lends credibility to the
energy wave function, taken on its own,
it does not constitute a conclusive proof that this provides the
best probability measure for local photon detection.

To reconsider the calculation of local photon detection rates, we consider a model detector
that consists of $N$ $\mathrm{He}^+$ ions in ion traps at locations $\bm{a}_I$.
The initial state has all the detector ions in their ground states and
includes a generic single-photon wave packet
$\phi_\alpha(\bm{k},t)=\phi_\alpha(\bm{k})\exp(-\,\mathrm{i}ckt)$,
\begin{equation}\label{eq:kpacket1}
\int\!d^3\bm{k}\,\sum_{\alpha=1}^2|\phi_\alpha(\bm{k})|^2=1,
\end{equation}
to describe the photon that eventually might be detected upon absorption.

Taking into account dipole selection rules, we find the scattering matrix
elements for photon absorption in the ion array,
\begin{eqnarray} \nonumber
  S_{fi}&=&\sqrt{\alpha_S c\omega_{n,1}}\int\!d^3\bm{k}\,\,\delta(\omega_{n,1}-ck)
  \sum_{I=1}^N\exp(\mathrm{i}\bm{k}\cdot\bm{a}_I)
  \\ \label{eq:Sfi1}
  &&\times
  \sum_{\alpha=1}^2\phi_\alpha(\bm{k})\bm{\epsilon}_\alpha(\bm{k})\cdot\!
  \sum_{m_1=-1}^1\!\langle n,1,m_1|\bm{\mathrm{y}}|1,0,0\rangle,
\end{eqnarray}
where the term with factor $\exp(\mathrm{i}\bm{k}\cdot\bm{a}_I)$ arises from excitation of
the $\mathrm{He}^+$ ion in the location $\bm{a}_I$.

Eq.~(\ref{eq:Sfi1}) is derived from first principles in Appendix B through integration
of the second-quantized Schr\"odinger equation for the ion array.
However, we can get the result (\ref{eq:Sfi1}) also from Eq.~(\ref{eq:matrix1})
through \\
$-$ substituting $|i\rangle=|1,0,0\rangle$ and $|f\rangle=|n,1,m_1\rangle$ for
the initial and final ion states, respectively;\\
$-$ taking into account that
dipole approximation amounts
to $\exp(\mathrm{i}\bm{k}\cdot\bm{\mathrm{y}})\to\exp(\mathrm{i}\bm{k}\cdot\bm{a}_I)$ for
the ion located at $\bm{y}=\bm{a}_I$;\\
$-$ summing and integrating the monochromatic result over polarization label $\alpha$
and photon wave vector $\bm{k}$ with
the wave packet components $\phi_\alpha(\bm{k})$.

We note that, without the energy preserving factor $\delta(\omega_{n,1}-ck)$,
the matrix element (\ref{eq:Sfi1}) would involve the Fourier transform of the wave packet
$\bm{\phi}(\bm{k})\equiv\sum_\alpha\phi_\alpha(\bm{k})\bm{\epsilon}_\alpha(\bm{k})$
at the positions $\bm{a}_I$, i.e.~the matrix element for photon absorption
in the ion trap at $\bm{a}_I$ would be directly
proportional to the Landau-Peierls wave function $\bm{\phi}(\bm{a}_I)$.
To infer more clearly what Eq.~(\ref{eq:Sfi1}) tells us about position-dependence
of photon absorption rates in the detector, we substitute,
\begin{equation}\label{eq:FTphi}
  \sum_{\alpha=1}^2\phi_\alpha(\bm{k})\bm{\epsilon}_\alpha(\bm{k})
  =\frac{1}{\sqrt{2\pi}^3}\int\!d^3\bm{x}\,
  \exp(-\,\mathrm{i}\bm{k}\cdot\bm{x})\bm{\phi}(\bm{x}),
\end{equation}
and perform the integration over $|\bm{k}|$. This yields
\begin{eqnarray} \nonumber
  S_{fi}
  &=&\sqrt{\frac{\alpha_S}{2\pi}}\left(\frac{\omega_{n,1}}{c}\right)^{5/2}
  \int_0^\pi\!d\theta\,\sin\theta\,
  \\ \nonumber
  &&\times\sum_{I=1}^N\int\!d^3\bm{x}\,
  \exp(\mathrm{i}\omega_{n,1}|\bm{a}_I-\bm{x}|\cos\theta/c)
  \\ \nonumber
  &&\times\bm{\phi}(\bm{x})\cdot
  \sum_{m_1=-1}^1\langle n,1,m_1|\bm{\mathrm{y}}|1,0,0\rangle
  \\ \nonumber
  &=&\sqrt{2\pi\alpha_S}\left(\frac{\omega_{n,1}}{c}\right)^{3/2}
  \,\sum_{I=1}^N\int\!d^3\bm{x}\,\frac{\sin(\omega_{n,1}|\bm{a}_I-\bm{x}|/c)}{\pi
    |\bm{a}_I-\bm{x}|}\,
  \\ \label{eq:Sfi2} 
  &&\times\bm{\phi}(\bm{x})\cdot
  \sum_{m_1=-1}^1\langle n,1,m_1|\bm{\mathrm{y}}|1,0,0\rangle.
\end{eqnarray}
This tells us that the photon absorption matrix element involves $N$ copies
of the convolution
of the Landau-Peierls wave function with the normalized sinc function,
evaluated at the detector sites,
\begin{equation}\label{eq:sincphi}
  (\mathrm{sinc}\circ\bm{\phi})(\bm{a}_I)=
  \int\!d^3\bm{x}\,\frac{\sin(\omega_{n,1}|\bm{a}_I-\bm{x}|/c)}{\pi
    |\bm{a}_I-\bm{x}|}\,\bm{\phi}(\bm{x}).
\end{equation}
Please note that this convolution arises solely as a consequence
of the Fourier transform relation (\ref{eq:FTphi}) between the $\bm{k}$-space
wave packet and the Landau-Peierls wave function. There is no requirement
nor assumption that the spatial integration in (\ref{eq:Sfi2},\ref{eq:sincphi})
arises from integration over a detector volume, nor is there a postulate
of spatial coarse graining involved to define a local operator.

The width of the sinc function at its base is $2\pi c/\omega_{n,1}=\lambda_n$.
We might infer from this that the Landau-Peierls wave function
provides a proxy for a probability density for photon location
through $|\bm{\phi}(\bm{x})|^2$: The magnitude
of $\bm{\phi}(\bm{x})$ in a region of
radius $\pi c/\omega_{n,1}=\lambda_{n}/2$ around $\bm{a}_I$ determines
the probability for photon detection by the ion trap in location $\bm{a}_I$,
\begin{eqnarray} \label{eq:Sfi3} 
  S_{fi}&\propto&\sum_{I=1}^N\langle\bm{\phi}(\bm{a}_I)\rangle_n\cdot
  \sum_{m_1=-1}^1\langle n,1,m_1|\bm{\mathrm{y}}|1,0,0\rangle.
\end{eqnarray}
Here we use the abbreviation
\begin{equation}
  \langle\bm{\phi}(\bm{a}_I)\rangle_n\equiv
  \int\!d^3\bm{x}\,\frac{\sin(\omega_{n,1}|\bm{a}_I-\bm{x}|/c)}{\pi
    |\bm{a}_I-\bm{x}|}\,\bm{\phi}(\bm{x})
\end{equation}
for the convolution (\ref{eq:sincphi})
to emphasize that the
position dependence of the absorption matrix element can be
expressed in terms of the volume integral of $\bm{\phi}(\bm{x})$
around $\bm{a}_I$ with the oscillating and decreasing sinc function.
We denote the calculation
of $\langle\bm{\phi}(\bm{a}_I)\rangle_n$ as ``local sampling''
of the function $\bm{\phi}(\bm{x})$ around the location $\bm{a}_I$.
Large $|\bm{\phi}(\bm{x})|^2$ in a region of
radius $\pi c/\omega_{n,1}=\lambda_{n}/2$ around $\bm{a}_I$ will
enhance the probability for photon detection in that region.

However, we will see that the same case can be made for the
energy wave function and the Hawton density:

We can use that the positive frequency part of the electric field
of the photon wave packet $\phi_\alpha(\bm{k})$ satisfies
\begin{eqnarray}\nonumber
  e\bm{E}^{(+)}(\bm{x},t)&=&\frac{\mathrm{i}\hbar}{2\pi}\sqrt{\alpha_Sc}
  \int\!d^3\bm{k}\,\sqrt{ck}
  \sum_{\alpha=1}^2\bm{\epsilon}_\alpha(\bm{k})\phi_\alpha(\bm{k})
  \\ \label{eq:Ephi}
  &&\times
  \exp[\mathrm{i}(\bm{k}\cdot\bm{x}-ckt)],
\end{eqnarray}
such that
  \begin{eqnarray}\nonumber
    \sqrt{ck}\sum_{\alpha=1}^2\bm{\epsilon}_\alpha(\bm{k})\phi_\alpha(\bm{k})
    &=&-\,\frac{\mathrm{i}e}{4\pi^2\hbar\sqrt{\alpha_Sc}}
    \\ \nonumber
    &&\times
    \int\!d^3\bm{x}\,\bm{E}^{(+)}(\bm{x})\exp(-\,\mathrm{i}\bm{k}\cdot\bm{x}).
\end{eqnarray}
  Therefore, we can write the scattering matrix element (\ref{eq:Sfi1}) also
  in the form
\begin{eqnarray} \nonumber
  S_{fi}&=&\frac{e}{4\pi^2\mathrm{i}\hbar}\int\!d^3\bm{x}\int\!d^3\bm{k}
  \sum_{I=1}^N
  \exp[\mathrm{i}\bm{k}\cdot(\bm{a}_I-\bm{x})]
  \\ \nonumber
  &&\times\delta(\omega_{n,1}-ck)\bm{E}^{(+)}(\bm{x})\cdot
  \sum_{m_1=-1}^1\langle n,1,m_1|\bm{\mathrm{y}}|1,0,0\rangle
  \\\nonumber
  &=&\frac{e\omega_{n,1}}{\mathrm{i}\hbar c^2}
  \,\sum_{I=1}^N\int\!d^3\bm{x}\,\frac{\sin(\omega_{n,1}|\bm{a}_I-\bm{x}|/c)}{\pi
    |\bm{a}_I-\bm{x}|}\,
  \\ \label{eq:Sfi1E}
  &&\times\bm{E}^{(+)}(\bm{x})\cdot
  \sum_{m_1=-1}^1\langle n,1,m_1|\bm{\mathrm{y}}|1,0,0\rangle,
\end{eqnarray}
which would also result from using the electric field
representation of the interaction picture Hamiltonian,
\begin{eqnarray}  
  \tilde{H}_D(t)
  &=&e\int\!d^3\bm{x}\sum_\sigma\psi_\sigma^+(\bm{x},t)
  \bm{x}\cdot\bm{E}(\bm{x},t)\psi_\sigma(\bm{x},t).
\end{eqnarray}
Eq.~(\ref{eq:Sfi1E}) implies that we can also write
\begin{eqnarray} \label{eq:Sfi3E} 
  S_{fi}&\propto&\sum_{I=1}^N\langle\bm{E}^{(+)}(\bm{a}_I)\rangle_n\cdot
  \sum_{m_1=-1}^1\langle n,1,m_1|\bm{\mathrm{y}}|1,0,0\rangle.
\end{eqnarray}
This result tells us that, with the same
caveat $\bm{E}^{(+)}(\bm{x})\to\langle\bm{E}^{(+)}(\bm{x})\rangle_n$
as for $\bm{\phi}(\bm{x})$,
we could just as well
accept Glauber's proposal to
use $\langle\Phi_i|\bm{\mathrm{E}}^{(-)}(\bm{x})\cdot\bm{\mathrm{E}}^{(+)}(\bm{x})|\Phi_i\rangle$,
equivalent to $|\bm{\Psi}(\bm{x})|^2$,
as a proxy for the probability of photon absorption in the position $\bm{x}$
from the state $|\Phi_i\rangle$.

Furthermore, in addition to Eq.~(\ref{eq:Ephi}), we can also use
\begin{eqnarray}\nonumber
  e\bm{A}^{(+)}(\bm{x},t)&=&\frac{\hbar}{2\pi}\sqrt{\alpha_S}
  \int\!d^3\bm{k}\,\frac{1}{\sqrt{k}}
  \sum_{\alpha=1}^2\bm{\epsilon}_\alpha(\bm{k})\phi_\alpha(\bm{k})
  \\ \label{eq:Aphi}
  &&\times
  \exp[\mathrm{i}(\bm{k}\cdot\bm{x}-ckt)]
\end{eqnarray}
to write
\begin{eqnarray} \nonumber
  S_{fi}&=&\frac{e\omega_{n,1}}{4\pi^2\hbar}\int\!d^3\bm{x}\int\!d^3\bm{k}
  \sum_{I=1}^N
  \exp[\mathrm{i}\bm{k}\cdot(\bm{a}_I-\bm{x})]
  \\ \nonumber
  &&\times\delta(\omega_{n,1}-ck)\bm{A}^{(+)}(\bm{x})\cdot
  \sum_{m_1=-1}^1\langle n,1,m_1|\bm{\mathrm{y}}|1,0,0\rangle
  \\\nonumber
  &=&\frac{e\omega^2_{n,1}}{\hbar c^2}
  \,\sum_{I=1}^N\int\!d^3\bm{x}\,\frac{\sin(\omega_{n,1}|\bm{a}_I-\bm{x}|/c)}{\pi
    |\bm{a}_I-\bm{x}|}\,
  \\ \label{eq:Sfi1A}
  &&\times\bm{A}^{(+)}(\bm{x})\cdot
  \sum_{m_1=-1}^1\langle n,1,m_1|\bm{\mathrm{y}}|1,0,0\rangle.
\end{eqnarray}
This implies with Eq.~(\ref{eq:Sfi1E}) that we can infer
\begin{eqnarray} \nonumber
  |S_{fi}|^2&\propto&\mathrm{i}\,\frac{\epsilon_0}{\hbar}\,
  \sum_{I=1}^N\left[
    \bm{d}_n^+\cdot\langle\bm{E}^{(-)}(\bm{a}_I)\rangle_n
    \langle\bm{A}^{(+)}(\bm{a}_I)\rangle_n\cdot\bm{d}_n\right.
    \\  \label{eq:SfiH}
    &&-\left.\bm{d}_n^+\cdot\langle\bm{A}^{(-)}(\bm{a}_I)\rangle_n
    \langle\bm{E}^{(+)}(\bm{a}_I)\rangle_n\cdot\bm{d}_n\right],
\end{eqnarray}
where
\begin{equation}
  \bm{d}_n=\sum_{m_1=-1}^1\langle n,1,m_1|\bm{\mathrm{y}}|1,0,0\rangle,
\end{equation}
and we used the fact that the photon can trigger a dipole transition in only one of
the ions to exclude the cross multiplication terms.

The observation (\ref{eq:SfiH})
adds credibility to the Hawton density (\ref{eq:rhoH}) for
the photon localization density, just as Eqs.~(\ref{eq:Sfi3}) and (\ref{eq:Sfi3E})
add credibility to the Landau-Peierls wave function or the energy wave function,
respectively.

We note that calculation of local detection rates, on its own, does not
unambiguously favor
one unique proposal for photon localization, and that the local absorption rates
do not directly yield the proposed photon localization probabilities, but involve
projections onto dipole matrix elements.

We also note that Eqs.~(\ref{eq:Sfi2},\ref{eq:Sfi1E}) imply sampling
of $\bm{\phi}(\bm{x})$ or $\bm{E}^{(+)}(\bm{x})$ over radii
of order $\lambda_n/2=11.4\,\mathrm{nm}\times n^2/(n^2-1)$,
although the Bohr radius for the photon-absorbing $\mathrm{He}^+$ ion is
only $a_0/2=0.265\,\mathrm{\AA}$. We can understand this as a consequence of Fourier
transformation. The $\mathrm{He}^+$ ion, as the basic photon-detecting unit, probes
$\bm{\phi}(\bm{k})$ through the excitation energy $E_{n,1}=hc/\lambda_n$, i.e.,
\textit{in the first place the ion probes $\bm{\phi}(\bm{k})$ on a sphere of
  radius $k_n=2\pi/\lambda_n$ in $\bm{k}$-space}. 
It is this primary probing in $\bm{k}$-space that then appears
like a probing of $\bm{\phi}(\bm{x})$ or $\bm{E}^{(+)}(\bm{x})$ 
over a volume $V\sim\pi\lambda_n^3/6$ in $\bm{x}$-space.

Mandel had argued within the framework
of his photon detection operator $\bm{a}(\bm{x},t)$, that spatial
coarse graining over volumes $V>\lambda^3$ is necessary to promote the 
operator $\mathcal{N}(\bm{x},t)=\bm{a}^+(\bm{x},t)\cdot\bm{a}(\bm{x},t)$
to an approximate local photon counting
operator $\mathcal{N}(V,t)=\int_V d^3\bm{x}\,\mathcal{N}(\bm{x},t)$,
\begin{equation}
  [\mathcal{N}(V,t),\bm{a}^+(\bm{x},t)]\simeq\left\{\begin{array}{l}
  \bm{a}^+(\bm{x},t)\,\,\mathrm{if}\,\,\bm{x}\in V,\\
  \,\,\,\,0\,\,\mathrm{otherwise}.\\
  \end{array}
  \right.
\end{equation}
 As a justification for the transition $\mathcal{N}(\bm{x},t)\to\mathcal{N}(V,t)$,
Mandel pointed out that a detector of area $S$ and sampling time $T$
probes a detection volume $cTS$ \cite{ref:mandel}. 

The present detector analysis
provides a different reason why detection of photons
with wavelength $\lambda$ appears to probe a volume of order $V\sim\lambda^3$,
with the sinc function as a weight factor, even when the volume
of the detecting element
is smaller than $\lambda^3$ (as is the case for ions or nanoAPDs).
It is first and foremost probing
of a sphere of radius $2\pi/\lambda=\Delta E/\hbar c$ in $\bm{k}$-space,
through an energy gap $\Delta E$ in the photon-absorbing detector element,
that translates through Fourier transformation into a signal that appears
to involve probing a volume of order $V\sim\lambda^3$ in $\bm{x}$-space.
We also note that this observation arises as a generic consequence
of the scattering matrix for photon absorption, and therefore is
not limited to using the operator $\mathcal{N}(\bm{x},t)$. 

\section{Comparison of the different proposals for photon position probability densities}
\label{sec:packet}

We discuss the different proposals for photon position probability densities for
a two-dimensional Gaussian isotropic single-photon wave packet
$\bm{\phi}(\bm{k},t)=\bm{\phi}(\bm{k})\exp(-\,\mathrm{i}ckt)$, 
\begin{equation}\label{eq:Phik1}
  \bm{\phi}(\bm{k})=\sqrt{\frac{2\Delta x^2}{\pi}}\exp\!\left(-\,\Delta x^2 k^2\right)
  \bm{\epsilon}(\bm{k}).
\end{equation}
The wave vector is $\bm{k}=k(\cos\vartheta,\sin\vartheta)$
and the corresponding polarization vector is
\begin{equation}
  \bm{\epsilon}(\bm{k})=\left(\begin{array}{r}
    -\,\sin\vartheta\\
    \cos\vartheta\\
  \end{array}
  \right).
\end{equation}
For convenience, we parametrize the photon wave packet in terms of
the width $\Delta x$ along a coordinate axis in two dimensions,
if either the vector Fourier transform (\ref{eq:LP2d}) or the scalar
Fourier transform (\ref{eq:FD2d}) of the wave packet is used to define
photon position uncertainty. The width of the packet in $k$ space is
$\Delta k=1/2\Delta x$.

The energy of the photon packet (\ref{eq:Phik1}) is
\begin{equation}
E=\hbar c\sqrt{\pi/\Delta x^2}.
\end{equation}

The Landau-Peierls wave function (\ref{eq:LP}) at the
position $\bm{x}=r(\cos\varphi,\sin\varphi)$ is
\begin{eqnarray}\nonumber
  \bm{\phi}(\bm{x},t)&=&\frac{1}{2\pi}\int\!d^2\bm{k}\,\exp(\mathrm{i}\bm{k}\cdot\bm{x})
  \bm{\phi}(\bm{k},t)
  \\ \nonumber
  &=&\sqrt{\frac{2\Delta x^2}{\pi}}\left(\begin{array}{r}
    -\,\sin\varphi\\
    \cos\varphi\\
  \end{array}
  \right)
  \\ \label{eq:LP2d}
  &&\times\!\int_0^\infty\!dk\,kJ_1(kr)\exp\!\left(-\,\Delta x^2 k^2-\mathrm{i}ckt\right)\!.
\end{eqnarray}
The Landau-Peierls wave function vanishes at the source position $r=0$
of the radially outgoing photon packet as a consequence of
\begin{eqnarray}\nonumber
  &&\int_0^{2\pi}\!d\vartheta\,\exp[\mathrm{i}kr\cos(\vartheta-\varphi)]
  \left(\begin{array}{r}
    -\,\sin\vartheta\\
    \cos\vartheta\\
  \end{array}
  \right)
  \\ \label{eq:AngAverage}
  &&\qquad=
  2\pi\mathrm{i} J_1(kr)\left(\begin{array}{r}
    -\,\sin\varphi\\
    \cos\varphi\\
  \end{array}
  \right)\!,
\end{eqnarray}
whereas the energy density $\mathcal{H}(\bm{x},0)$ has its maximum there.
This has nothing to do with the nonlocal relation between the Landau-Peierls
wave function and the electromagnetic fields of the wave packet (\ref{eq:Phik1}),
but is merely a consequence of averaging over directions in the origin.
We will see that for the same reason (\ref{eq:AngAverage}), the energy
wave function $\bm{\Psi}(\bm{x},t)$
and the Hawton density $\varrho_H(\bm{x},t)$ also vanish at $r=0$.

There is only one polarization
component $\phi(\bm{k},t)=|\bm{\phi}(\bm{k})|\exp(-\,\mathrm{i}ckt)$ in two dimensions,
and the proposal (\ref{eq:FD}) yields
\begin{eqnarray}\nonumber
  \psi(\bm{x},t)&=&\frac{1}{2\pi}\int\!d^2\bm{k}\,\exp(\mathrm{i}\bm{k}\cdot\bm{x})
  \phi(\bm{k},t)=\sqrt{\frac{2\Delta x^2}{\pi}}
  \\  \label{eq:FD2d}
  &&\times\!\int_0^\infty\!dk\,kJ_0(kr)
  \exp\!\left(-\,\Delta x^2 k^2-\mathrm{i}ckt\right)\!.
\end{eqnarray}
The absence of averaging over directions yields a maximum for $|\psi(\bm{x},0)|^2$
at $r=0$. We find the same behavior for $\mathcal{H}(\bm{x},0)$ due to the magnetic
contribution $B^{(\pm)}(\bm{x},0)$, which also is not affected by
Eq.~(\ref{eq:AngAverage}).

The positive frequency components of the vector potential and
the electromagnetic fields in two dimensions for
the wave packet $\bm{\phi}(\bm{k},t)$ are
\begin{eqnarray}\nonumber
  &&\bm{A}^{(+)}(\bm{x},t)
  =\mathrm{i}\sqrt{\frac{\hbar\Delta x^2}{\pi\epsilon_0 c}}\left(\begin{array}{r}
    -\,\sin\varphi\\
    \cos\varphi\\
  \end{array}
  \right)
  \\ \label{eq:A2d}
  &&\qquad\times\!\int_0^\infty\!dk\,\sqrt{k}
  J_1(kr)\exp\!\left(-\,\Delta x^2 k^2-\mathrm{i}ckt\right)\!,
\end{eqnarray}
\begin{eqnarray}\nonumber
  &&\frac{1}{\mathrm{i}}\sqrt{\epsilon_0}\bm{E}^{(+)}(\bm{x},t)
  =\bm{\Psi}(\bm{x},t)/\sqrt{2}
  \\ \nonumber
  &&\quad=\mathrm{i}\sqrt{\frac{\hbar c\Delta x^2}{\pi}}\left(\begin{array}{r}
    -\,\sin\varphi\\
    \cos\varphi\\
  \end{array}
  \right)
  \\ \label{eq:Psi2d}
  &&\qquad\times\!\int_0^\infty\!dk\,\sqrt{k}^3
  J_1(kr)\exp\!\left(-\,\Delta x^2 k^2-\mathrm{i}ckt\right)\!,
\end{eqnarray}
and
\begin{eqnarray}\nonumber
  &&\frac{1}{\mathrm{i}\sqrt{\mu_0}}B^{(+)}(\bm{x},t)
  =\sqrt{\frac{\hbar c\Delta x^2}{\pi}}
  \\
  &&\qquad\times\!\int_0^\infty\!dk\,\sqrt{k}^3
  J_0(kr)\exp\!\left(-\,\Delta x^2 k^2-\mathrm{i}ckt\right)\!.
\end{eqnarray}
Note that we have in
two dimensions\footnote{The corresponding
relation in three dimensions is
$\bm{B}(\bm{x},t)/\sqrt{\mu_0}\simeq\langle\hat{\bm{k}}\rangle\times
\sqrt{\epsilon_0}\bm{E}(\bm{x},t)$.}
\begin{equation}\label{eq:BRrel2d}
  B(\bm{x},t)/\sqrt{\mu_0}\simeq\sqrt{\epsilon_0}\hat{\bm{z}}\cdot[\langle\hat{\bm{k}}\rangle\times
  \bm{E}(\bm{x},t)]
\end{equation}
(with $\hat{\bm{z}}\cdot[\langle\hat{\bm{k}}\rangle\times\bm{E}(\bm{x},t)]$
defined through embedding the two dimensions as an $(x,y)$ plane into three dimensions)
only for narrow wave packets centered
around the expectation value $\langle\bm{k}\rangle$, $\Delta k\ll|\langle\bm{k}\rangle|$.
However, this does not hold for an isotropic wave packet.

The relation between the energy wave function and the Landau-Peierls wave function
in two dimensions is
\begin{equation}\label{eq:Psitophi2d}
  \bm{\phi}(\bm{x},t)=\frac{1}{\pi\sqrt{2\hbar c}}\frac{\Gamma[3/4]}{\Gamma[1/4]}
  \int\!d^2\bm{x}'\,\frac{\bm{\Psi}(\bm{x}',t)}{|\bm{x}-\bm{x}'|^{3/2}}.
\end{equation}
Formal inversion in two dimensions motivates again a relation
involving a singular kernel,
\begin{equation}\label{eq:phitoPsi2d}
  \bm{\Psi}(\bm{x},t)\sim\frac{\sqrt{\hbar c}}{8\pi\sqrt{2}}\frac{\Gamma[1/4]}{\Gamma[3/4]}
  \int\!d^2\bm{x}'\,\frac{\bm{\phi}(\bm{x}',t)}{|\bm{x}-\bm{x}'|^{5/2}},
\end{equation}
with the same caveats that apply to the relation (\ref{eq:phitoPsi3d}).

Cook's wave function in two dimensions has components
\begin{equation}
  \bm{\phi}_1(\bm{x},t)=\bm{\phi}(\bm{x},t)/\sqrt{2},\quad
  \phi_2(\bm{x},t)=\psi(\bm{x},t)/\sqrt{2},
\end{equation}
such that the corresponding position probability density is the arithmetic mean
of the Landau-Peierls density $|\bm{\phi}(\bm{x},t)|^2$ and $|\psi(\bm{x},t)|^2$.
Therefore, we only plot the densities $|\bm{\phi}(\bm{x},t)|^2$, $|\psi(\bm{x},t)|^2$,
$|\bm{\Psi}(\bm{x},t)|^2/E$, $\mathcal{H}(\bm{x},t)/E$, and $\varrho_H(\bm{x},t)$.

\begin{figure}[hbt]
\scalebox{0.9}{\includegraphics{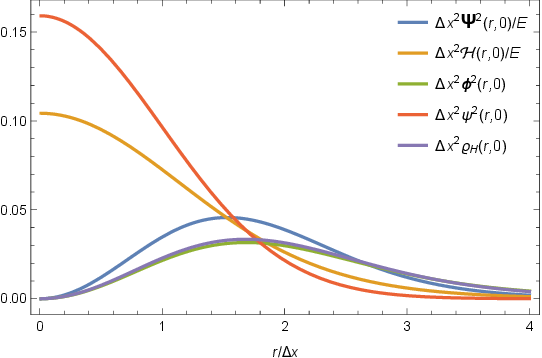}}
\caption{\label{fig:All5t0}
  The proposals $|\bm{\Psi}(\bm{x},t)|^2/E$, $\mathcal{H}(\bm{x},t)/E$,
  $|\bm{\phi}(\bm{x},t)|^2$, 
  $|\bm{\psi}(\bm{x},t)|^2$ and $\varrho_H(\bm{x},t)$ for $t=0$.}
\end{figure}

\begin{figure}[hbt]
\scalebox{0.9}{\includegraphics{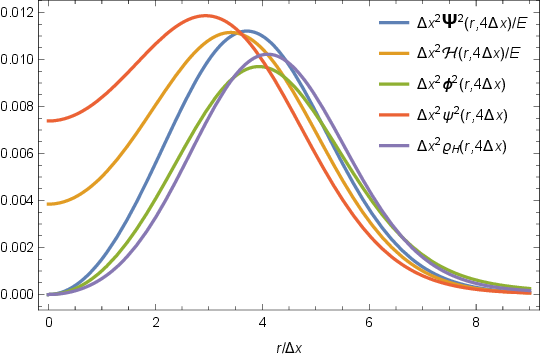}}
\caption{\label{fig:All5t4}
  The proposals $|\bm{\Psi}(\bm{x},t)|^2/E$, $\mathcal{H}(\bm{x},t)/E$,
  $|\bm{\phi}(\bm{x},t)|^2$, 
  $|\bm{\psi}(\bm{x},t)|^2$ and $\varrho_H(\bm{x},t)$ for $ct=4\Delta x$.}
\end{figure}

\begin{figure}[hbt]
\scalebox{0.9}{\includegraphics{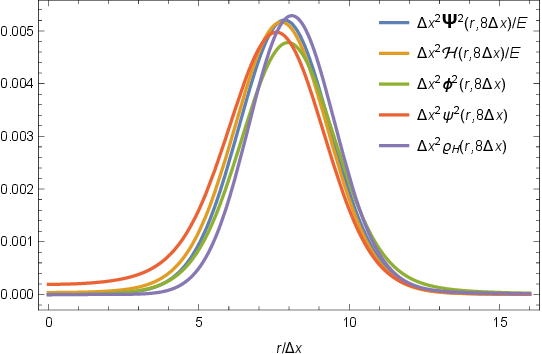}}
\caption{\label{fig:All5t8}
  The proposals $|\bm{\Psi}(\bm{x},t)|^2/E$, $\mathcal{H}(\bm{x},t)/E$,
  $|\bm{\phi}(\bm{x},t)|^2$, 
  $|\bm{\psi}(\bm{x},t)|^2$ and $\varrho_H(\bm{x},t)$ for $ct=8\Delta x$.}
\end{figure}

\begin{figure}[hbt]
\scalebox{0.9}{\includegraphics{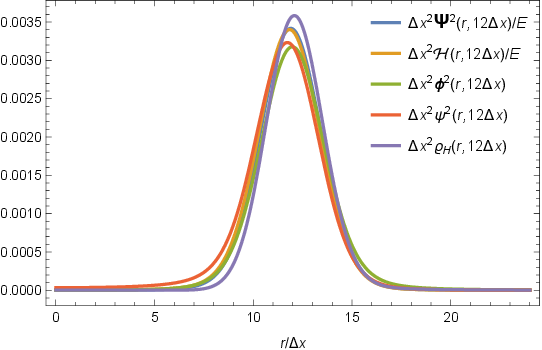}}
\caption{\label{fig:All5t12}
  The proposals $|\bm{\Psi}(\bm{x},t)|^2/E$, $\mathcal{H}(\bm{x},t)/E$,
  $|\bm{\phi}(\bm{x},t)|^2$, 
  $|\bm{\psi}(\bm{x},t)|^2$ and $\varrho_H(\bm{x},t)$ for $ct=12\Delta x$.}
\end{figure}

The Hawton density (\ref{eq:rhoH}) for the two-dimensional wave
packet (\ref{eq:Phik1}),
\begin{eqnarray}\nonumber
  \varrho_H(\bm{x},t)&=&\frac{\Delta x^2}{\pi}\int_0^\infty\!dk\int_0^\infty\!dq\,kq
  \left(\sqrt{k/q}+\sqrt{q/k}\right)J_1(kr)
  \\ 
  &&\times J_1(qr)\exp[-\,\Delta x^2(k^2+q^2)\cos[(k-q)ct],
\end{eqnarray}
differs from the Landau-Peierls density $|\bm{\phi}(\bm{x},t)|^2$ (\ref{eq:LP2d}),
if expressed as a double integral over wave numbers $k$ and $q$, through
the factor $(\sqrt{k/q}+\sqrt{q/k})/2$. It differs from the corresponding
expression for $|\bm{\Psi}(\bm{x},t)|^2/E$
(\ref{eq:Psi2d}) through a factor $\sqrt{\pi/\Delta x^2}(q^{-1}+k^{-1})/2$.
However, the wave number integrals are exponentially cut off at the upper end
at $1/\Delta x=\Delta k/2$, and they are cut off linearly at the lower end through
the factor $kq$. This means that the wave number integrals defining the
three probability densities $|\bm{\phi}(\bm{x},t)|^2$, $|\bm{\Psi}(\bm{x},t)|^2/E$
and $\varrho_H(\bm{x},t)$ are concentrated to a
region around $k\sim q\lesssim 1/\Delta x$. This explain why those three
proposals for photon probability densities do not strongly deviate from
each other, as depicted in Figs.~\ref{fig:All5t0}-\ref{fig:All5t12}.

All five proposals for a photon position probability density trace the classically
expected spherical wave front $r=ct$ for distances and times $r\simeq ct\gg \Delta x$.
We can understand this from the asymptotic behavior of the Bessel functions entering
into the calculations of the proposals for photon probability densities.
The wave number integrals are exponentially cut off at $k\sim 1/\Delta x$.
This implies that we can consider the asymptotic forms of the Bessel functions for
$r\gg\Delta x$:
\begin{equation*}
  J_0(kr)\to\frac{1}{\sqrt{2\pi kr}}\left[\frac{1-\mathrm{i}}{\sqrt{2}}\exp(\mathrm{i}kr)
    +\frac{1+\mathrm{i}}{\sqrt{2}}\exp(-\,\mathrm{i}kr)\right],
\end{equation*}
\begin{equation*}
  J_1(kr)\to-\,\frac{1}{\sqrt{2\pi kr}}\left[\frac{1+\mathrm{i}}{\sqrt{2}}\exp(\mathrm{i}kr)
    +\frac{1-\mathrm{i}}{\sqrt{2}}\exp(-\,\mathrm{i}kr)\right].
\end{equation*}
This implies that the wave number integrals at late times
are dominated by the substitutions
\begin{eqnarray}\nonumber
  &&J_n(kr)\exp(-\Delta x^2 k^2-\mathrm{i}ckt)
  \\ \nonumber
  &&\to (-)^n\,\frac{\exp[(-)^{n+1}\pi/4]}{\sqrt{2\pi kr}}\exp(-\Delta x^2 k^2)
  \exp[\mathrm{i}k(r-ct)],
\end{eqnarray}
which favor contributions from the region $r\simeq ct$ from avoidance of destructive interference.
The terms proportional to $\exp[-\,\mathrm{i}k(r+ct)]$, on the other hand, will always suffer
destructive interference and are therefore suppressed in the probability densities.
Therefore, all the proposed photon probability densities coalesce around the classical
spherical wave front $r=ct$ at late times.

\begin{figure}[htb]
\scalebox{0.9}{\includegraphics{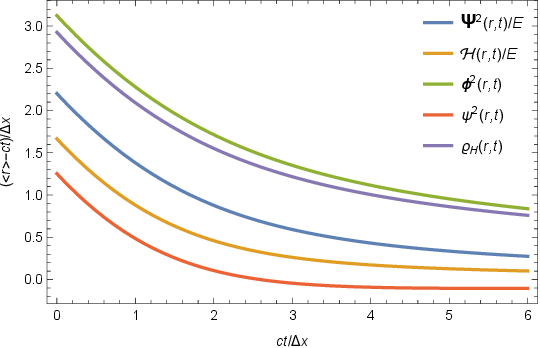}}
\caption{\label{fig:rMINUSct}
  The deviation $\langle r\rangle-ct$ for the different proposed photon
  localization measures for $0\le ct\le 6\Delta x$.}
\end{figure}

The finite widths in position space at $t=0$ imply that all photon localization
measures will start out with nonvanishing expectation values $\langle r\rangle|_{t=0}>0$,
i.e.~they will all evolve with $\langle r\rangle-ct>0$ and coalesce towards $\langle r\rangle-ct=0$.
This is displayed in Fig.~\ref{fig:rMINUSct}.

The Landau-Peierls wave function $\bm{\phi}(\bm{x},t)$
converges slowest towards $\langle r\rangle-ct=0$ while the scalar Fourier transform
$\psi(\bm{x},t)$ reaches $\langle r\rangle-ct\simeq 0$ already after a time $ct\simeq 4\Delta x$
has elapsed. The coalescence of the different position space wave packets for late times also
implies that their widths approach the width $\Delta x=1/2\Delta k$ of the vector 
Fourier transform $\bm{\phi}(\bm{x},t)$ and the scalar Fourier transform $\psi(\bm{x},t)$
after sufficiently many time steps $\Delta t=(c\Delta k)^{-1}$ have elapsed.

\section{Conclusions}\label{sec:conc}

The analysis of the different proposals for a measure of photon location in the framework
of an isotropic Gaussian single-photon packet of width $\Delta k$ showed that all measures coalesce
into an outgoing spherical wave front $\langle r\rangle=ct$
after sufficiently many time steps $(c\Delta k)^{-1}$ have elapsed.
However, the
scalar Fourier transform $\psi(\bm{x},t)$ and the normal ordered energy
density $\mathcal{H}(\bm{x},t)$ indicate a maximum of photon location at time $t=0$
in the position $\bm{x}=0$, where the isotropic packet starts out into all directions.
This is very different from the behavior of the other proposals.
The Landau-Peierls wave function $\bm{\phi}(\bm{x},t)$, the energy wave function
$\bm{\Psi}(\bm{x},t)$, and the Hawton density $\varrho_H(\bm{x},t)$ start out with maxima
at values $r>0$ (Fig.~\ref{fig:All5t0}) and vanish in $r=0$, as a consequence of angular
integration over polarization of the isotropic wave packet.

We also found that local photon absorption rates could be expressed through local
sampling
of the Landau-Peierls wave function, the energy wave function, or a tensor version of
the Hawton density. All those three proposals for local photon probability densities
also remain close throughout the time evolution of the single-photon
wave packet. This indicates that those three propositions are preferred for providing
probability measures for photon location.

We summarize findings in Table \ref{tab:tools}. The results for
Cook's two-vector wave function can be
inferred from $\bm{\phi}_1(\bm{x},t)=\bm{\phi}(\bm{x},t)/\sqrt{2}$,
while $\bm{\phi}_2(\bm{x},t)$ does not appear in the local photon detection
rates in Sec.~\ref{sec:detect}.

\begin{table}[ht]
\begin{center}
\begin{tabular}{|c|l|l|}\hline
Tool & Detection & Local\\
\hline
$\bm{\phi}(\bm{x},t)$ & Yes & No\\
\hline
$\left(\begin{array}{c}
  \bm{\phi}_1(\bm{x},t)\\
  \bm{\phi}_2(\bm{x},t)\\
\end{array}
\right)$ & No & No\\
\hline
$\mathcal{H}(\bm{x},t)$ & No & Yes\\
\hline
$\bm{\Psi}(\bm{x},t)$ & Yes & Yes\\
\hline
$\varrho_H(\bm{x},t)$ & Yes & No\\
\hline
$\psi_\alpha(\bm{x},t)$ & No & No\\
\hline
\end{tabular}
\caption{Comparison of different tools that have been proposed for
  the description of local photon densities.
  ``Detection'' refers to whether local photon absorption rates can be
  expressed in terms of the proposed tool through local sampling,
  as described in Sec.~\ref{sec:detect}. ``Local'' refers to whether
the tool can locally be expressed in terms of electric or magnetic fields.}
\label{tab:tools}
\end{center}
\end{table}

We recall from Sec.~\ref{sec:detect} that proposed photon localization measures
which can be inferred from photon absorption matrix elements, do
not require recourse to a derivation through spatial coarse graining for
them to appear as products of a negative frequency term times a
positive frequency term. That product structure is implied by virtue
of the photon absorption matrix elements. However, the
requirement of excitation across an energy gap $\Delta E$ in the detecting
element implies probing the single-photon wave packet $\bm{\phi}(\bm{k})$ over
a sphere $k=\Delta E/\hbar c$ in momentum space. This translates into
probing local photon position amplitudes $\bm{\phi}(\bm{x})$
or $\bm{E}^{(+)}(\bm{x})$ across volumes $V\sim\lambda^3$ in $\bm{x}$-space,
even when the detecting device is much smaller than $V$.

We have noted the many nonlocal relations between the different proposals
for the description of photon probability densities. However, there
are instances when we do care about locality, or at least quasilocality,
when selecting a wave function or probability density for photons.
On the one hand, we consider local relation with the electromagnetic
fields $\bm{E}^{(+)}(\bm{x},t)$ or $\bm{B}^{(+)}(\bm{x},t)$ as highly desirable,
both for local interaction between photons and charged particles and
for gauge invariance. On the other hand, our requirement
to define position probability density through a local detection probability
density, implies that the proposed probability measure should have a close
relation with the corresponding transition matrix elements.
We found that this latter requirement cannot be satisfied locally. However,
it can be satisfied in a quasilocal manner through local sampling
of the wave function around the detector site. We found that
the Landau-Peierls wave function, the energy wave function, and
the Hawton density satisfy this requirement.

We note that the preference of several leading experts
in the field is to use the energy wave function as a probability density for
photon location. This is also in line with the practice
in optics to use $\bm{E}^2(\bm{x},t)$ as a measure for ``intensity'' of light, and it is
supported by our observation that the energy wave function is the only tool that meets
both of the desirable requirements of direct relation
to local photon detection probabilities, as expressed through the scattering matrix,
and of local expression in terms of electromagnetic fields.
Our observations therefore reaffirm the energy wave function (\ref{eq:Psi1})
as the most promising tool for the description of local photon densities.

We also note that the reasoning that led us
to $\bm{\Psi}(\bm{x},t)=-\,\mathrm{i}\sqrt{2\epsilon_0}\bm{E}^{(+)}(\bm{x},t)$,
to describe the local photon probability density
through $|\bm{\Psi}(\bm{x},t)|^2/E$, was based on the observation that photon
counting generically occurs through electric dipole transitions.

If photon detectors would not be dominated by electric dipole transitions,
but by magnetic spin flips through the Pauli term, detector analysis would
lead us to a magnetic energy wave function
$\bm{\Psi}_B(\bm{x},t)=-\,\mathrm{i}\sqrt{2/\mu_0}\bm{B}^{(+)}(\bm{x},t)$
and a photon probability density $|\bm{\Psi}_B(\bm{x},t)|^2/E$.

Our suggestion that local position probability density only makes sense
as a proxy for local detection probability then implies
a contextual nature of wave functions and position probability
densities.

\section*{Appendix A: Convolution representations}

To derive the various convolution relations between the electric field,
the Landau-Peierls wave function, the vector potential, and the Hawton density,
we start with the rescaled energy wave function,
  \begin{eqnarray} \nonumber
    \bm{\psi}(\bm{x},t)&=&\frac{1}{\sqrt{2\pi}^3}\int\!d^3\bm{k}\,\sqrt{k}
    \bm{\phi}(\bm{k})\exp[\mathrm{i}(\bm{k}\cdot\bm{x}-ckt)]
    \\ \label{eq:A+2}
    &=&\frac{\bm{\Psi}(\bm{x},t)}{\sqrt{\hbar c}}
    =-\,\mathrm{i}\sqrt{\frac{2\epsilon_0}{\hbar c}}\bm{E}^{(+)}(\bm{x},t),
  \end{eqnarray}
  and define
  \begin{eqnarray} \nonumber 
    \bm{\psi}_\nu(\bm{x},t)
    &=&\frac{1}{\sqrt{2\pi}^3}\int\!d^3\bm{k}\,k^{\nu+\frac{1}{2}}
    \bm{\phi}(\bm{k})\exp[\mathrm{i}(\bm{k}\cdot\bm{x}-ckt)]
    \\ \nonumber
    &=&\int\!d^3\bm{x}'\,D_\nu(\bm{x}-\bm{x}')\bm{\psi}(\bm{x}',t)
    \\ \label{eq:Anu1}
    &=&\int\!d^3\bm{x}'\,D_{\nu+1/2}(\bm{x}-\bm{x}')\bm{\phi}(\bm{x}',t).
  \end{eqnarray}
  We have $D_0(\bm{x})=\delta(\bm{x})$ and for $-\,3<\nu<0$ and $\epsilon\to 0$,
\begin{eqnarray} \nonumber
D_\nu(\bm{x})&=&\frac{1}{8\pi^3}\int\!d^3\bm{k}\,k^{\nu}
\exp(\mathrm{i}\bm{k}\cdot\bm{x}-\epsilon k)
\\ \nonumber
&=&\frac{1}{4\pi^2 \mathrm{i}r}\int_0^\infty\!dk\,k^{\nu+1}\Big(\exp[-\,(\epsilon-\mathrm{i}r)k]
\\ \nonumber
&&-\,\exp[-\,(\epsilon+\mathrm{i}r)k]\Big)
\\ \nonumber
&=&\frac{\Gamma(\nu+2)}{4\pi^2 \mathrm{i}r}\left[
\frac{1}{(\epsilon-\mathrm{i}r)^{\nu+2}}-\frac{1}{(\epsilon+\mathrm{i}r)^{\nu+2}}
\right]
\\ \label{eq:Dnu3}
&=&\frac{\Gamma(\nu+2)}{2\pi^2 r^{\nu+3}}\sin\!\left(\pi\frac{\nu+2}{2}\right).
\end{eqnarray}
This yields in particular
  \begin{eqnarray} \nonumber
    \bm{\psi}_{-1/2}(\bm{x},t)&=&\bm{\phi}(\bm{x},t)
    \\ \label{eq:psi2phi}
    &=&\frac{1}{4\sqrt{2\pi^3}}\int\!d^3\bm{x}'\,
    \frac{\bm{\psi}(\bm{x}',t)}{|\bm{x}-\bm{x}'|^{5/2}},
    \end{eqnarray}
  and
  \begin{eqnarray} \nonumber
    \bm{\psi}_{-1}(\bm{x},t)&=&\sqrt{\frac{2\epsilon_0 c}{\hbar}}\,\bm{A}^{(+)}(\bm{x},t)
    \\ \nonumber
    &=&\frac{1}{4\sqrt{2\pi^3}}\int\!d^3\bm{x}'\,
    \frac{\bm{\phi}(\bm{x}',t)}{|\bm{x}-\bm{x}'|^{5/2}}
    \\ \label{eq:Psi2A}
    &=&\frac{1}{2\pi^2}\int\!d^3\bm{x}'\,
    \frac{\bm{\psi}(\bm{x}',t)}{|\bm{x}-\bm{x}'|^2}.
  \end{eqnarray}

  The formal inversion (\ref{eq:phitoPsi3d}) of Eq.~(\ref{eq:psi2phi}), which has been used in
  Refs.~\cite{ref:MW,ref:amrein}, involves
  the kernel $D_{1/2}(\bm{x})$. In a similar vein, formal inversion
  of the relation (\ref{eq:Psi2A}) through a kernel $D_1(\bm{x})$ leads to a singular mapping 
  $\bm{A}^{(+)}(\bm{x},t)\to\bm{E}^{(+)}(\bm{x},t)$,
   \begin{eqnarray} \label{eq:A2Psi}
    \bm{\psi}(\bm{x},t)&\sim&-\,\frac{1}{\pi^2}\int\!d^3\bm{x}'\,
    \frac{\bm{\psi}_{-1}(\bm{x}',t)}{|\bm{x}-\bm{x}'|^4}.
  \end{eqnarray}
   Bia{\l}ynicki-Birula's reasoning in favor of $\bm{\Psi}(\bm{x},t)$
   over $\bm{\phi}(\bm{x},t)$ from the lack of an actual inversion
   of Eq.~(\ref{eq:psi2phi}) then also favors $\bm{\Psi}(\bm{x},t)$
   over $\bm{A}^{(+)}(\bm{x},t)$ or $\varrho_H(\bm{x},t)$.

   Eq.~(\ref{eq:Psi2A}) also implies the relation
\begin{eqnarray} \nonumber 
\varrho_H(\bm{x},t)&=&\int\!d^3\bm{x}'\,
\frac{1}{4\pi^2|\bm{x}-\bm{x}'|^2}
\\ \nonumber
&&\times\left[\bm{\psi}^+(\bm{x},t)\cdot\bm{\psi}(\bm{x}',t)
  +\bm{\psi}^+(\bm{x}',t)\cdot\bm{\psi}(\bm{x},t)\right]
\\\nonumber
&=&\int\!d^3\bm{x}'\,
\frac{1}{2\pi^2|\bm{x}-\bm{x}'|^2}
\\ \nonumber
&&\times\left[\mathrm{Re}\,\bm{\psi}(\bm{x},t)\cdot\mathrm{Re}\,\bm{\psi}(\bm{x}',t)\right.
  \\ \label{eq:rhoHpsi2}%
  &&\quad+\left.\mathrm{Im}\,\bm{\psi}(\bm{x},t)\cdot\mathrm{Im}\,\bm{\psi}(\bm{x}',t)\right].
\end{eqnarray}

Although the Hawton density is not manifestly positive definite,
the representation (\ref{eq:rhoHpsi2}) suggests that it will be positive definite
for every energy wave function with single maxima for
$[\mathrm{Re}\,\bm{\psi}(\bm{x},t)]^2$ and $[\mathrm{Im}\,\bm{\psi}(\bm{x},t)]^2$.

The convolution $[D_{-1/2}\circ D_{-1/2}](\bm{x})=D_{-1}(\bm{x})$
then proves again that the Hawton density is normalized if and only if
the Landau-Peierls wave function is normalized,
\begin{eqnarray} \nonumber
\int\!d^3\bm{x}\,\varrho_H(\bm{x},t)&=&\int\!d^3\bm{x}\int\!d^3\bm{x}'\,
\frac{\bm{\psi}^+(\bm{x},t)\cdot\bm{\psi}(\bm{x}',t)}{2\pi^2|\bm{x}-\bm{x}'|^2}
\\
&=&\int\!d^3\bm{x}\,|\bm{\phi}(\bm{x},t)|^2.
\end{eqnarray}
This also follows directly by expressing $\bm{A}^{(+)}(\bm{x},t)$
and $\bm{E}^{(+)}(\bm{x},t)$ through the $\bm{k}$-space wave packet
$\bm{\phi}(\bm{k},t)$ \cite{ref:HawtonMelde}.

The convolution kernels $D_\nu(\bm{x})$, which relate the vectors
 $\bm{E}^{(+)}(\bm{x},t)$, $\bm{\phi}(\bm{x},t)$, and $\bm{A}^{(+)}(\bm{x},t)$,
also relate the corresponding  helicity based
functions $E_\alpha^{(+)}(\bm{x},t)$, $\psi_\alpha(\bm{x},t)$, and $A_\alpha^{(+)}(\bm{x},t)$.

The two-dimensional version of the kernel
(\ref{eq:Dnu3}) for $-\,2<\nu<-\,1/2$ is given by,
\begin{eqnarray} \nonumber
D^{(2)}_\nu(\bm{x})&=&\frac{1}{4\pi^2}\int\!d^2\bm{k}\,k^{\nu}
\exp(\mathrm{i}\bm{k}\cdot\bm{x})
\\ \nonumber
&=&\frac{1}{2\pi}\int_0^\infty\!dk\,k^{\nu+1}J_0(kr)
\\ \label{eq:Dnu2}
&=&\frac{2^\nu}{\pi r^{\nu+2}}\frac{\Gamma[1+\nu/2]}{\Gamma[-\,\nu/2]}.
\end{eqnarray}

\section*{Appendix B: Matrix elements for photon absorption in an array of ion traps}

We are interested in the question what absorption of
a general single-photon state $|\phi(t)\rangle$ (\ref{eq:LPmandel},\ref{eq:gamma2})
with wave
packet $\bm{\phi}(\bm{k})=\sum_\alpha\phi_\alpha(\bm{k})\bm{\epsilon}_\alpha(\bm{k})$
tells us about absorption probabilities in $\bm{x}$-space.
Here, we will develop the
corresponding first-order scattering matrix elements directly from integration
of the second-quantized Schr\"odinger equation. We will use minimal photon
coupling.

The wave function for the initial state $|\Phi_i(t')\rangle$ with $N$ $\mathrm{He}^+$ ions
in their ground states and the photon state
has the $2^{N+1}$ components
\begin{eqnarray} \nonumber
  &&\langle\{\bm{y},s\};\bm{k},\alpha|\Phi_i(t')\rangle
  =
  \phi_\alpha(\bm{k})\exp\!\left[-\,\mathrm{i}(NE_1+\hbar ck)t'/\hbar\right]
  \\ \nonumber
  &&\qquad\times\prod_{I=1}^N\phi_{1,0,0}(\bm{y}_I-\bm{a}_I)\chi_{s_I}
  \\ \label{eq:initPhi}
  &&\quad=\phi_\alpha(\bm{k})
  \exp\!\left[-\,\mathrm{i}(NE_1+\hbar ck)t'/\hbar\right]
  \langle\{\bm{y},s\}|\tilde{\Phi}_i\rangle,
\end{eqnarray}
\begin{equation}
  \sum_{s_I=-\,1/2}^{1/2}\left|\chi_{s_I}\right|^2=1,
\end{equation}
corresponding to the electron and photon polarizations. 
Here, $\{\bm{y},s\}=\bm{y}_1,s_1;\ldots;\bm{y}_N,s_N$ denotes the coordinates and spin projections
of the $N$ electrons. We suppress the electron spin labels
$s_I$ and weight factors $\chi_{s_I}$
in the following since the dominant photon interactions do not change spin polarizations.

The Hamiltonian of the system is in leading order of the photon couplings
(in the Schr\"odinger picture)
\begin{eqnarray} \nonumber
  H&=&K+V-\int\!d^3\bm{x}\,\bm{j}(\bm{x})\cdot\bm{A}(\bm{x})
  \\
  &=&H_0-\int\!d^3\bm{x}\,\bm{j}(\bm{x})\cdot\bm{A}(\bm{x}),
\end{eqnarray}
with the kinetic terms for the helium nuclei and the electrons
and photons\footnote{We neglect any possible
presence of ${}^3\mathrm{He}$ isotopes. In this appendix,
we follow the standard convention to
use $\psi_\sigma(\bm{x},t)$ for $\bm{x}$-space electron 
operators with spin projection $\hbar\sigma=\pm\,\hbar/2$.
Please do not confuse with the photon amplitudes $\psi_\alpha(\bm{x},t)$
in the polarization basis, as defined in Eq.~(\ref{eq:mandelmandel2})
and used in Secs.~\ref{sec:intro} and \ref{sec:packet}.}
\begin{eqnarray}\nonumber
  K&=&\int\!d^3\bm{x}\,\Bigg(\frac{\hbar^2}{2M}\bm{\nabla}\Psi^+(\bm{x})\cdot\bm{\nabla}\Psi(\bm{x})
  \\ \nonumber
  &&+\left.\frac{\hbar^2}{2m}\sum_{\sigma=-1/2}^{1/2}
  \bm{\nabla}\psi_\sigma^+(\bm{x})\cdot\bm{\nabla}\psi_\sigma(\bm{x})\right)
  \\
  &&+\int\!d^3\bm{k}\sum_{\alpha=1}^2\hbar ck \,a_\alpha^+(\bm{k})a_\alpha(\bm{k}),
\end{eqnarray}
the Coulomb terms
\begin{eqnarray}\nonumber
  V&=&\int\!d^3\bm{x}\int\!d^3\bm{x}'\,\frac{e^2}{8\pi\epsilon_0|\bm{x}-\bm{x}'|}
  \\ \nonumber
  &&\times
  \Bigg(4\Psi^+(\bm{x})\Psi^+(\bm{x}')\Psi(\bm{x}')\Psi(\bm{x})
  \\ \nonumber
  &&+\sum_{\sigma,\sigma'=-1/2}^{1/2}\psi_\sigma^+(\bm{x})\psi_{\sigma'}^+(\bm{x}')\psi_{\sigma'}(\bm{x}')
  \psi_\sigma(\bm{x})
  \\
  &&-\left.4\sum_{\sigma=-1/2}^{1/2}\psi_\sigma^+(\bm{x})\Psi^+(\bm{x}')\Psi(\bm{x}')\psi_\sigma(\bm{x})
  \right),
\end{eqnarray}
the electric current density
\begin{eqnarray}\nonumber
  \bm{j}(\bm{x})&=&\frac{e\hbar}{\mathrm{i}M}\left[\Psi^+(\bm{x})\cdot\bm{\nabla}\Psi(\bm{x})
    -\bm{\nabla}\Psi^+(\bm{x})\cdot\Psi(\bm{x})\right]
  \\ \nonumber
  &&-\,\frac{e\hbar}{2\mathrm{i}m}\sum_{\sigma=-1/2}^{1/2}\left[
    \psi_\sigma^+(\bm{x})\cdot\bm{\nabla}\psi_\sigma(\bm{x})\right.
    \\ \label{eq:current1}%
    &&-\left.\bm{\nabla}\psi_\sigma^+(\bm{x})\cdot\psi_\sigma(\bm{x})\right],
\end{eqnarray}
and the photon operator
\begin{eqnarray}\nonumber
\bm{A}(\bm{x})&=&\sqrt{\frac{\hbar\mu_0 c}{(2\pi)^3}}
\int\!\frac{d^3\bm{k}}{\sqrt{2k}}\sum_{\alpha=1}^2
\bm{\epsilon}_\alpha(\bm{k})\left[
a_\alpha(\bm{k})
\exp(\mathrm{i}\bm{k}\cdot\bm{x})\right.
\\ \label{eq:Amode}
&&
+\left.a_{\alpha}^+(\bm{k})
\exp(-\,\mathrm{i}\bm{k}\cdot\bm{x})\right].
\end{eqnarray}

Since the photon coupling to the electrons is $M/2m\simeq 3.65\times 10^3$ stronger than to
the He nuclei, we neglect the He nuclei. We took that already into account in the
wave function (\ref{eq:initPhi}) by assigning only one position variable $\bm{y}_I$ to each
atom, which we take as the electron position. We consider the He nuclei fixed in the
positions $\bm{a}_I$. We also limit the discussion to the
detection of photons with energies below the ionization energy 54.4 eV of $\mathrm{He}^+$,
such that photon absorption will only excite the $\mathrm{He}^+$ ions. Absorption of the
photon in leading order will then generate excited final states with wave functions 
\begin{eqnarray}\nonumber
  &&\langle\bm{y}_1,\ldots,\bm{y}_N|\Phi^{(n,\ell,m_\ell)}_J(t)\rangle
  =\exp\!\left(-\,\mathrm{i}[(N-1)E_1+E_n]t/\hbar\right)
  \\ \label{eq:finalPhi} %
  &&\qquad\times\,
  \phi_{n,\ell,m_\ell}(\bm{y}_J-\bm{a}_J)\prod_{I=1,I\neq J}^N
  \phi_{1,0,0}(\bm{y}_I-\bm{a}_I),
\end{eqnarray}
if the photon wave packet contains wave vectors such that the condition
$ck=\omega_{n,1}=(E_n-E_1)/\hbar$ can be fulfilled.

Specifically, the evolution of the initial state (\ref{eq:initPhi}) in the limit $t'\to -\,\infty$
to the excited state (\ref{eq:finalPhi}) in the limit $t\to\infty$ is governed
by the integration of the Schr\"odinger equation,
\begin{eqnarray}\nonumber
  |\Phi_f\rangle&=&\exp(\mathrm{i}H_0t/\hbar)\exp[-\,\mathrm{i}H(t-t')/\hbar]
  \\ \nonumber
  &&\times\,\exp(-\,\mathrm{i}H_0t'/\hbar)|\Phi_i\rangle
  \\
  &=&\mathrm{T}\exp\!\left[-\,\frac{\mathrm{i}}{\hbar}
    \int_{t'}^t\!d\tau\,H_D(\tau)\right]|\Phi_i\rangle,
\end{eqnarray}
where the Dirac (interaction)
picture Hamiltonian on the states contains the free quantum fields
\begin{equation*}
  \psi_\sigma(\bm{x},t)=\frac{1}{\sqrt{2\pi}^3}\int\!d^3\bm{k}\,c_\sigma(\bm{k})
  \exp\!\left[\mathrm{i}\left(\bm{k}\cdot\bm{x}
    -\frac{\hbar\bm{k}^2}{2m}t\right)\right]
\end{equation*}
and
\begin{eqnarray}\nonumber
\bm{A}(\bm{x},t)&=&\sqrt{\frac{\hbar\mu_0 c}{(2\pi)^3}}
\int\!\frac{d^3\bm{k}}{\sqrt{2k}}\sum_{\alpha=1}^2
\bm{\epsilon}_\alpha(\bm{k})
\\ \nonumber
&&\times\,\Big(
a_\alpha(\bm{k})
\exp[\mathrm{i}(\bm{k}\cdot\bm{x}-ckt)]
\\ \label{eq:Amodet}
&&
+\,a_{\alpha}^+(\bm{k})
\exp[-\,\mathrm{i}(\bm{k}\cdot\bm{x}-ckt)]\Big)
\end{eqnarray}
of the interaction picture,
\begin{eqnarray}\nonumber
  H_D(t)
  &=&\int\!d^3\bm{x}\,\frac{e\hbar}{2\mathrm{i}m}\sum_\sigma
  \left[\psi_\sigma^+(\bm{x},t)\cdot\bm{\nabla}\psi_\sigma(\bm{x},t)\right.
    \\
    &&-\left.
    \bm{\nabla}\psi_\sigma^+(\bm{x},t)\cdot\psi_\sigma(\bm{x},t)\right]\cdot\bm{A}(\bm{x},t).
\end{eqnarray}
This yields in leading order 
\begin{eqnarray} \nonumber
  |\Phi_f\rangle&=&|\Phi_i\rangle-\int_{-\,\infty}^\infty\!dt\,\exp(\mathrm{i}H_0t/\hbar)
  \int\!d^3\bm{x}\,\frac{e}{2m}\bm{A}(\bm{x})
  \\ \nonumber
  &&\times\sum_\sigma
  \left[\psi_\sigma^+(\bm{x})\cdot\bm{\nabla}\psi_\sigma(\bm{x})
    -\bm{\nabla}\psi_\sigma^+(\bm{x})\cdot\psi_\sigma(\bm{x})\right]|\Phi_i\rangle
  \\ \nonumber
  &&\times\,\exp\!\left[-\,\mathrm{i}(NE_1+\hbar ck)t/\hbar\right]
  \\ \nonumber
  &=&|\Phi_i\rangle-\frac{e\hbar}{4m}\int\!d^3\bm{k}\,\sqrt{\frac{\hbar\mu_0 c}{\pi k}}\,
  \delta(H_0-NE_1-\hbar ck)
  \\ \nonumber
  &&\times
  \int\!d^3\bm{x}\,\exp(\mathrm{i}\bm{k}\cdot\bm{x})
  \sum_{\sigma,\alpha}\phi_\alpha(\bm{k})\bm{\epsilon}_\alpha(\bm{k})
  \\ \nonumber
  &&\times
  \left[\psi_\sigma^+(\bm{x})\cdot\bm{\nabla}\psi_\sigma(\bm{x})
    -\bm{\nabla}\psi_\sigma^+(\bm{x})\cdot\psi_\sigma(\bm{x})\right]|\tilde{\Phi}_i\rangle.
\end{eqnarray}
Evaluation of the electron operators yields after projection onto electron coordinates
(with $\{\bm{y}\}=\bm{y}_1,\ldots,\bm{y}_N$),
\begin{eqnarray} \nonumber
  \langle\{\bm{y}\}|\Phi_f\rangle&=&
  \langle\{\bm{y}\}|\Phi_i\rangle-
  \frac{e\hbar}{2m}\int\!d^3\bm{k}\,\sqrt{\frac{\hbar\mu_0 c}{\pi k}}
  \\ \nonumber
  &&\times
  \delta(h_0-NE_1-\hbar ck)\sum_{\alpha=1}^2\phi_\alpha(\bm{k})\bm{\epsilon}_\alpha(\bm{k})
  \\ \label{eq:Swavefunction}
  &&\times 
  \sum_{J=1}^N\exp(\mathrm{i}\bm{k}\cdot\bm{y}_J)\frac{\partial}{\partial\bm{y}_J}
  \langle\{\bm{y}\}|\tilde{\Phi}_i\rangle.
\end{eqnarray}
The operator $h_0$ is the first-quantized representation of $H_0$ with respect to the
electronic degrees of freedom (recall that we neglect the nuclear degrees of freedom
as the helium nuclei are assumed fixed a the locations $\bm{a}_I$),
\begin{eqnarray} \nonumber
  h_0&=&-\sum_{I=1}^N\frac{\hbar^2}{2m}\frac{\partial^2}{\partial\bm{y}_I^2}
  -\sum_{I=1}^{N}\sum_{J=1}^{N}\frac{e^2}{2\pi\epsilon_0|\bm{y}_I-\bm{a}_J|}
  \\ \nonumber
  &&+\sum_{I=1}^{N-1}\sum_{J=I+1}^N\frac{e^2}{4\pi\epsilon_0|\bm{y}_I-\bm{y}_J|}
  \\
  &&+\int\!d^3\bm{k}\sum_{\alpha=1}^2\hbar ck \,a_\alpha^+(\bm{k})a_\alpha(\bm{k}).
\end{eqnarray}
This arises from the basic formula for transition from second to first quantization,
\begin{eqnarray} \nonumber
  &&\langle\bm{y}|\int\!d^3\bm{x}\left(\frac{\hbar^2}{2m}\frac{\partial\psi^+(\bm{x})}{
    \partial\bm{x}}\cdot\frac{\partial\psi(\bm{x})}{\partial\bm{x}}+\psi^+(\bm{x})V(\bm{x})\psi(\bm{x})
  \right)
  \\ \nonumber
  &&=\langle 0|\psi(\bm{y})\int\!d^3\bm{x}\bigg(\frac{\hbar^2}{2m}\frac{\partial\psi^+(\bm{x})}{
    \partial\bm{x}}\cdot\frac{\partial\psi(\bm{x})}{\partial\bm{x}}
  \\
  &&\quad+\,\psi^+(\bm{x})V(\bm{x})\psi(\bm{x})
  \bigg)
  =\left(-\,\frac{\hbar^2}{2m}\frac{\partial^2}{\partial\bm{y}^2}+V(\bm{y})\right)
  \!\langle\bm{y}|.
\end{eqnarray}
In the following, we neglect interatomic interactions,
\begin{eqnarray} \nonumber
  h_0&\to&-\sum_{I=1}^N\frac{\hbar^2}{2m}\frac{\partial^2}{\partial\bm{y}_I^2}
  -\sum_{I=1}^{N}\frac{e^2}{2\pi\epsilon_0|\bm{y}_I-\bm{a}_I|}
  \\
  &&+\int\!d^3\bm{k}\sum_{\alpha=1}^2\hbar ck \,a_\alpha^+(\bm{k})a_\alpha(\bm{k}),
\end{eqnarray}
such that we can use unperturbed electron wave functions
$\phi_{n,\ell,m_\ell}(\bm{y}_I-\bm{a}_I)$ for the detector ions.
Eq.~(\ref{eq:Swavefunction}) is then a superposition of a single-photon state
\begin{eqnarray} \nonumber
  \langle\bm{y}_1,\ldots,\bm{y}_N|\Phi_i\rangle&=&
  \sum_{\alpha=1}^2\int\!d^3\bm{k}\, a_\alpha^+(\bm{k})|0\rangle
  \phi_\alpha(\bm{k})
  \\
  &&\times\prod_{I=1}^N\phi_{1,0,0}(\bm{y}_I-\bm{a}_I)
\end{eqnarray}
with derivatives of the zero-photon state
\begin{eqnarray} 
  \langle\bm{y}_1,\ldots,\bm{y}_N|\tilde{\Phi}_i\rangle&=&|0\rangle
  \prod_{I=1}^N
  \phi_{1,0,0}(\bm{y}_I-\bm{a}_I).
\end{eqnarray}

For the projection into the zero-photon sector through projection onto all
possible detector states, we can use dipole approximation,
$\exp(\mathrm{i}\bm{k}\cdot\bm{y}_J)\simeq\exp(\mathrm{i}\bm{k}\cdot\bm{a}_J)$,
because the photon wavelength satisfies $\lambda>22.8$ nm
and the relevant Bohr radius for the $\mathrm{He}^+$ ions is $a_0/2=0.265$ \AA.

We also switch to length form,
\begin{eqnarray} \nonumber
  \int\!d^3\bm{y}\,\langle n,\ell,m_\ell|\bm{y}\rangle\frac{\partial}{\partial\bm{y}}
  \langle\bm{y}|1,0,0\rangle
  &=&
  -\,\langle n,\ell,m_\ell|\bm{\mathrm{y}}|1,0,0\rangle
  \\ \label{eq:p2x} %
  &&\times
  \frac{m}{\hbar^2}(E_n-E_1),
\end{eqnarray}
where we denote the position operator for position $\bm{y}$ in upright notation.

Taking into account dipole selection rules, we find that the possibly nonvanishing
projections of $\langle\bm{y}_1,\ldots,\bm{y}_N|\Phi_f\rangle$ into the zero-photon sector
are the corresponding scattering matrix elements
\begin{eqnarray} \nonumber
  S_{fi}&=&\sqrt{\alpha_S c\omega_{n,1}}\int\!d^3\bm{k}\,\,\delta(\omega_{n,1}-ck)
  \sum_{I=1}^N\exp(\mathrm{i}\bm{k}\cdot\bm{a}_I)
  \\ \label{eq:Sfi1b}
  &&\times
  \sum_{\alpha=1}^2\phi_\alpha(\bm{k})\bm{\epsilon}_\alpha(\bm{k})\cdot\!
  \sum_{m_1=-1}^1\!\langle n,1,m_1|\bm{\mathrm{y}}|1,0,0\rangle,
\end{eqnarray}
where the term with factor  $\exp(\mathrm{i}\bm{k}\cdot\bm{a}_I)$ arises from excitation of
the $\mathrm{He}^+$ ion in the location $\bm{a}_I$.

Of course, the dipole matrix element $\langle n,1,m_1|\bm{\mathrm{y}}|1,0,0\rangle$ has
the same value for each photon-absorbing detector ion and therefore does not carry
the label $I$ anymore. Mathematically, the dependence on $I$ disappears because
\begin{equation}
  \langle n,1,m_1|(\bm{\mathrm{y}}-\bm{a}_I)|1,0,0\rangle
  =\langle n,1,m_1|\bm{\mathrm{y}}|1,0,0\rangle.
\end{equation}


\section*{Acknowledgments}

RD thanks Margaret Hawton for discussions and for clarifications on Ref.~\cite{ref:Hawton2}.
We acknowledge support from the Natural Sciences and Engineering Research Council of Canada
(NSERC) and from MITACS.

\end{document}